# Chlorine and Bromine Isotope Fractionation of Halogenated Organic Pollutants on Gas Chromatography Columns

**Caiming Tang[1,2], Jianhua Tan[3], Songsong Xiong[1,2], Jun Liu[1,2], Yujuan Fan[1,2], Xianzhi Peng[1,*]**

*[1] State Key Laboratory of Organic Geochemistry, Guangzhou Institute of Geochemistry, Chinese Academy of Sciences, Guangzhou 510640, China*

*[2] University of Chinese Academy of Sciences, Beijing 100049, China*

*[3] Guangzhou Quality Supervision and Testing Institute, Guangzhou, 510110, China*

*Corresponding Author.

Tel: +86-020-85290009; Fax: +86-020-85290009. E-mail: pengx@gig.ac.cn. (X. Peng).

## ABSTRACT

Compound-specific chlorine/bromine isotope analysis (CSIA-Cl/Br) has become a useful approach for degradation pathway investigation and source appointment of halogenated organic pollutants (HOPs). CSIA-Cl/Br is usually conducted by gas chromatography–mass spectrometry (GC–MS) techniques, which could be negatively impacted by chlorine and bromine isotope fractionation of HOPs on GC columns. In this study, 31 organochlorines and 4 organobromines were systematically investigated in terms of chlorine/bromine isotope fractionation on GC columns using GC–double focus magnetic-sector high resolution MS (GC–DFS-HRMS). On-column chlorine/bromine isotope fractionation behaviors of the HOPs were explored, presenting various isotope fractionation modes and extents. Twenty-nine HOPs exhibited inverse isotope fractionation, and only polychlorinated biphenyl-138 (PCB-138) and PCB-153 presented normal isotope fractionation. And no observable isotope fractionation was found for the rest four HOPs, i.e., PCB-101, 1,2,3,7,8-pentachlorodibenzofuran, PCB-180 and 2,3,7,8-tetrachlorodibenzofuran. The isotope fractionation extents of different HOPs varied from below the observable threshold (5.0‰) to 73.1‰ (PCB-18). The mechanisms of the on-column chlorine/bromine isotope fractionation were tentatively interpreted with the Craig-Gordon model and a modified two-film model. Inverse isotope effects and normal isotope effects might contribute to the total isotope effects together and thus determine the isotope fractionation directions and extents. Proposals derived from the main results of this study for CSIA-Cl/Br research were provided for improving the precision and accuracy of CSIA-Cl/Br results. The findings of this study will shed light on the development of CSIA-Cl/Br methods using GC–MS techniques, and help to implement the research using CSIA-Cl/Br to investigate the environmental behaviors and pollution sources of HOPs.

## INTRODUCTION

Gas chromatography (GC) is a powerful analytical tool and has been widely used in a variety of areas such as environmental analysis, petrochemical industry, product quality inspection, bioanalysis, and metabolomics.[1] Due to its high-efficient separation capability, GC was even applied to the separation of isotope-labeled compounds from their natives.[2,3] GC has been proved to be capable of separating isotopologues of various compounds, such as caffeine and caffeine-1,3,7-$(CD_3)_3$,[4] *n*-alkanes ($C_{15}$-$C_{17}$) and their perdeuterated isotopomers,[5] as well as formamide derivatives and their deuterated isotopomers.[6] Most of these reported studies involved the separation of hydrogen and carbon isotopologues.[7-9] In some cases, although the isotopologues could not be sufficiently separated, the discrepancies between the retention times of the isotopologues were evidently distinguishable.[8] Separation of isotopologues on GC columns is a form of isotope fractionation which is a process affecting the relative abundances of isotopes. Isotope fractionation can be caused by both chemical reactions (e.g., synthesis and decomposition) and physical changes (such as volatilization and diffusion).[10-12]

Isotope fractionation occurring on GC columns is a physical change process, of which the theories and mechanisms were studied and proposed previously.[3] The substitution of lighter isotope with the heavier can cause molecules becoming more hydrophobic, which thus always leads to inverse isotope effect on GC columns.[3] This inverse isotope effect, in other words, means that the heavier isotopologues elute faster than the lighter on GC columns. Bermejo et al. investigated the separation of chlorinated mixtures of dimethylbenzene and dimethylbenzene-$d_{10}$ on four GC columns coated with different stationary phases, and found that eight pairs of isotopomeric chlorinated products exhibited inverse isotope effect.[13] Matucha et al. studied the isotope-specific retention behaviors of alkane isotopomers on GC columns, and found that the inverse isotope effect on the retention times of the alkane isotopomers increased with increase of the carbon length or the deuterated degree, and with decrease of column temperature.[9] They concluded that the inverse isotope effect on GC column was because of that the interaction between the analytes and stationary phase was dominated by van der Waals dispersion forces. Shi et al. separated nine isotopomeric pairs (hydrogen/deuterium) of molecules and quantitatively determined these isotopomers using GC with a DB-5 column, and found all pairs of isotopomeric molecules presented inverse isotope effects.[14] The results of the study indicated that the isotope effects were combinable, and the authors proposed that the GC separation processes had two substeps, i.e., mixing of analytes with liquid stationary phase, and condensation-vaporization of analytes. The former step was relatively isotope-insensitive, while the latter contributed dominantly to the observed inverse isotope effects, due to that intermolecular van der Waals forces are effective in the condensed phase and thus leads to changes of isotope-sensitive zero point energy when a molecule is condensed from gas phase. Up to now, however, the data about halogen (Cl/Br) isotope fractionation of halogenated organic pollutants (HOPs) on GC columns are extremely scarce.

GC coupled with mass spectrometry techniques such as isotope ratio MS (IRMS),[15,16] quadrupole MS (qMS),[16-20] ICP-MS,[21] and hybrid quadrupole time-of-flight MS (Q-TOF-MS)[22] have been applied to compound specific isotope analysis (CSIA). CSIA has become a mature analytical approach in several areas, such as food authenticity test, doping control in sports, and environmental research.[23-25] In environmental studies, CSIA has been applied to reaction pathway probing and source appointment for environmental pollutants.[23,26,27] Up to now, CISA studies of several elements including hydrogen, carbon, oxygen, nitrogen, sulfur, chlorine and bromine have been reported.[28,29] Recently, CISA-Cl/Br has become an emerging highlight and a pretty challenging task in environmental sciences.[28,30] Multi-dimensional CSIA of two or more elements has been used for in-depth revealing the environmental behaviors of pollutants.[31-33] Most of the reported CSIA studies employed GC and/or GC-related techniques for separation of investigated compounds. Accordingly, GC plays an important role in CSIA studies, particularly for the online (continuous-flow) CSIA research. Thus, the separation performance of GC may impact CSIA results. This separation performance is not only with regard to the separation of the investigated compounds from others (such as interferences), but also the unwanted intermolecular isotope fractionation of the compounds of interest in GC system. It has been reported that the carbon isotope fractionation occurring on the GC column or the whole GC system would cause deviations for carbon CSIA results.[34,35] Holmstrand et al. reported a normal chlorine isotope effect of chlorine isotopologues of 1,1,1-trichloro-2,2-bis(pchlorophenyl)ethane (DDT) on a preparative megabore-column capillary gas chromatography (pcGC).[36] They suggested that partial collection of DDT eluting from pcGC led to biased results in offline IRMS detection due to the isotope fractionation on the pcGC column. Presently, no available publication has reported the chlorine isotope fractionation of organochlorides on analytical GC column in online (continuous-flow) GC–MS (including IRMS, qMS, ICP-MS and Q-TOF-MS) system. Moreover, no study has revealed the bromine isotope fractionation of organobromine compounds on any chromatographic column or system. Nevertheless, chlorine/bromine isotope fractionation on GC columns could trigger deviations to the CSIA-Cl/Br results of organohalogen compounds. Therefore, chlorine/bromine isotope fractionation of chlorinated/brominated organic compounds on GC columns needs to be systematically and sufficiently investigated and ascertained.

In this study, we used a variety of HOPs including organochlorides and organobromines to ascertain Cl/Br isotope fractionation on analytical GC columns by GC–double focus magnetic-sector high resolution MS (GC–DFS-HRMS). Chlorine/bromine isotope fractionation behaviors of the HOPs on GC columns were revealed with the developed CSIA-Cl/Br method. The results obtained in this study will shed light on the method development of CSIA-Cl/Br for HOPs using GC–MS techniques, and benefit the studies using CSIA-Cl/Br to explore the environmental behaviors and contamination sources of HOPs.

## EXPERIMENTAL SECTION

Contents about *Chemicals*, *Materials* and *Preparation of solutions* were provided in the Supporting Information.

**Instrumental Analysis.** GC–HRMS consisted of dual Trace-GC-Ultra gas chromatographers coupled with a DFS-HRMS and a TriPlus autosampler (GC–DFS-HRMS, Thermo-Fisher Scientific, Bremen, Germany).

The prepared working solutions were directly analyzed by GC-HRMS. Two GC columns were used, i.e., a DB-5MS capillary column (60 m × 0.25 mm, 0.25 μm thickness) and a DB-5MS capillary column (30 m × 0.25 mm, 0.1 μm thickness, J&W Scientific, USA). The two columns were installed in the dual gas chromatographers, respectively and applied to separating different categories of compounds with different physicochemical properties such as chromatographic retention factor and thermal stability. In addition, GC temperature programs varied for separation of different categories of compounds. The details of columns and temperature programs for analyzing all the HOPs are provided in Table S-1.

The working conditions and parameters of MS are documented as follows: ionization was performed with a positive electronic impact (EI+) ionization source; electron impact energy was set at 45 eV; temperature of ionization source was set at 250 $^{o}$C; filament current was 0.8 mA; multiple ion detection (MID) mode was applied; dwell time for each isotopologue ion was $20 \pm 2$ ms; mass resolution (5% peak-valley definition) was tuned to ≥ 10000 and the MS detection accuracy was set at ±0.001 u; HRMS was calibrated in real time during MID operation with either perfluorotributylamine or perfluorokerosene.

Chemical structures of the investigated compounds were drawn with ChemDraw (Ultra 7.0, Cambridgesoft), and the exact masses of the molecular isotopologues were calculated with mass accuracy of 0.00001 u. Only the Cl/Br isotopologues were taken into account, which means the isotopologues containing D/T, $^{13}C$, and/or $^{17}O$/$^{18}O$ were not chosen (except $^{13}C_6$-HBB, for which only $^{13}C$ was taken into account). For a compound containing *n* Cl or Br atoms, all its molecular isotopologues (*n+1*) were chosen. By subtracting the mass of an electron from the calculated exact mass of each isotopologue, the mass-to-charge ratio (*m/z*) of the isotopologue molecular ion can be obtained. The *m/z* values were imported into MID methods to monitor the investigated compounds. The information with respect to the isotopologues of the investigated compounds, such as retention times, isotopologue chemical formulas, exact masses and exact *m/z* values, are provided in Table S-2.

**Data Processing.** The isotope ratio (R) was calculated as:

$$R = \frac{\sum_{i=0}^{n} i \cdot S_i}{\sum_{i=0}^{n} (n-i) \cdot S_i} \qquad (1)$$

where *n* is the number of Cl or Br atoms of a molecule; *i* is the number of $^{37}Cl$ or $^{81}Br$ atoms in an isotopologue; $S_i$ is the MS signal intensity of the molecular isotopologue *i*.

As shown in Figure 1, a chromatographic peak in the total ion chromatogram (TIC) was equally divided into 5 segments based on its retention time range (except DDTs and p,p'-DDD, of which the peaks were divided into three or four segments due to relatively low signal intensity and constancy). In every segment, the average MS signal intensities of all the isotopologues of each investigated compound were exported, and the isotope ratio can thus be calculated with Equation 1. The overall isotope ratio was calculated with the average MS signal intensities of all the isotopologues extracted from the whole chromatographic peak. All the exported MS signal intensity data were subjected to background subtraction. Data from 5 replicated injections were used to calculate the mean value and standard deviation (SD, 1σ) of isotope ratio.

Relative variations of isotope ratios derived from different retention-time segments of the HOPs were calculated with Equation 2:

$$\Delta^{h}\mathrm{E} = \left( \frac{R_{Tj}}{R_{overall}} - 1 \right) \cdot 1000‰ \qquad (2)$$

where $\Delta^{h}E$ is the relative variation of isotope ratio in each retention-time segment referenced to the overall isotope ratio of the corresponding compound; $R_{Tj}$ is the isotope ratio derived from the j’th retention-time segment ($T_j$); $R_{overall}$ is the overall isotope ratio.

The isotope fractionation extent ($\Delta'^{h}E$) was calculated with Equation 3:

$$\Delta'^{h}\mathrm{E} = \left( \frac{R_{T\text{-}first}}{R_{T\text{-}last}} - 1 \right) \cdot 1000‰ \qquad (3)$$

where $R_{T\text{-}first}$ is the average isotope ratio derived from the first retention-time segments (T1) of five replicated injections; $R_{T\text{-}last}$ is the average isotope ratio derived from the last retention-time segments of five replicated injections.

## RESULTS AND DISCUSSION

Contents about the performances of CSIA-Cl/Br method developed in this study are provided in the Supporting Information.

**Evaluation Schemes for On-column Chlorine/Bromine Isotope Fractionation.** For evaluation of on-column chlorine/bromine isotope fractionation, we divided each chromatographic peak into several equal segments in terms of the retention time range, and then calculated the individual isotope ratio derived from reach retention-time segment. Besides isotope ratios, $\Delta^{h}E$ ($\Delta^{37}Cl$ and $\Delta^{81}Br$) values calculated with Equation 2 were applied to further clearly illustrating the on column isotope fractionation. In addition, the isotope fractionation extents were evaluated with $\Delta'^{h}E$ values calculated by Equation 3. By means of these evaluation methods, the chlorine/bromine isotope fractionation of HOPs on GC columns were visually and explicitly elucidated (Figure 1).

**On-column Isotope Fractionation Modes.** As shown in Figure 2 (G1-G4) and Table S-3, the chlorine isotope ratios declined from the first retention-time segment to the last (T1-T5) for most of the investigated organochlorines including all PCDDs, DDTs, DDDs and DDEs. This indicated that the heavier isotopologues of these 25 chlorinated compounds eluted ahead to the lighter ones. The heavier isotopologues thus enriched in the front retention-time segments and decreased in the hind segments. On the contrary, the lighter isotopologues reduced in the front retention-time segments and enriched in the tail segments. Therefore, inverse kinetics of isotope effects of these compounds presented; In other words, inverse chlorine isotope fractionation of these compounds took place. Statistically insignificant chlorine isotope fractionation was observed for four chlorinated compounds, i.e., PCB-101, Penta-CDF-1, PCB-180 and TCDF (Figure 2, G5). The chlorine isotope ratios of these four compounds varied insignificantly, and no evident changing trend was observed. Thus, no chlorine isotope fractionation of these compounds on GC columns were found. Only PCB-138 and PCB-153 exhibited normal isotope fractionation on GC columns. The chlorine isotope ratios of these two compounds increased along with retention-time segments from T1 to T5 (Figure 2, G6). Contrary to most of the other investigated organochlorine compounds, the heavier isotopologues of PCB-138 and PCB-153 “run” slower than the lighter on GC columns.

As Figure 3 illustrates, the positive $\Delta^{37}$Cl values (above the dashed zero lines) indicates the enrichment of heavier isotope of chlorine, and the negative values (below the zero lines) demonstrates the decrease of $^{37}$Cl. The absolute values of $\Delta^{37}$Cl reflects the extents of isotope fractionation of the chlorinated compounds in the corresponding retention-time segments. Most of the investigated chlorinated compounds presented positive $\Delta^{37}$Cl values in the first two retention-time segments, and had negative values in the last two segments (Figure 3, G1-G3). This indicated the heavier isotope ($^{37}$Cl) enriched in the front part of chromatographic peaks and decreased in the tail compartment. And the inverse isotope fractionation thus can be observed. In Figure 3 (G4), the $\Delta^{37}$Cl values of PCB-101, Penta-CDF-1, PCB-180 and TCDF were close to the zero line and no observable changing trend was found. Therefore, these four compounds could be considered to have no isotope fractionation on GC columns. And in Figure 3 (G5), normal on-column isotope fractionation of PCB-138 and PCB-153 can be deduced.

It is notable that the intersection points of the plotted lines and the zero lines were much closed to the $\Delta^{37}$Cl values derived from the middle retention-time segments. This indicates that the isotope ratios derived from the middle retention-time segments were most equal to the overall isotope ratios of the corresponding compounds. At the middle segments, enrichment and decrease of the heavier isotope were balanced.

Four brominated compounds, BDE-77, $^{13}C_6$-HBB, HBB and OBDD, were investigated in terms of bromine isotope fractionation on GC columns. As can be concluded from Figure 2 (G7) and Figure 3 (G6), all the four brominated compounds exhibited inverse bromine isotope fractionation on GC columns. The detailed results about on-column bromine isotope fractionation of the four organobromines are documented in Table S-3.

**Extents of On-column Isotope Fractionation.** The extents of chlorine/bromine isotope fractionation of the investigated HOPs can be expressed as $\Delta'^{h}E$ values (Table S-4). In this study, if the absolute $\Delta'^{h}E$ values were higher than 10‰, within 5‰-10‰, or lower than 5‰, then the corresponding HOPs were considered to have significant, slight, or none isotope fractionation on GC columns, respectively. As shown in Figure 4, 30 compounds exhibited significant isotope fractionation, accounting for 85.7% of all the HOPs. Only one compound (PCB-138) showed slight isotope fractionation with the $\Delta'^{37}Cl$ value of -6.9‰. And four compounds, i.e., PCB-101, PCB-180, TCDF and Penta-CDF-1, presented none isotope fractionation, showing $\Delta'^{37}Cl$ values from -3.8‰ to 3.1‰.

PCB-18 presented the most significant on-column isotope fractionation, with the highest $\Delta'^{37}Cl$ value of 73.1‰. While its isomer PCB-28 exhibited evidently lower isotope fractionation, of which the $\Delta'^{37}Cl$ value was 46.5‰. A clear declining tendency was observed for the $\Delta'^{37}Cl$ value of PCBs with the increase of substituted Cl atoms. From PCB-18 (Tri-PCB) to PCB-101 (Penta-PCB), the corresponding $\Delta'^{37}Cl$ values were significantly reduced from 73.1‰ to 3.1‰. OCDD had the second highest extent of isotope fractionation, presenting $\Delta'^{37}Cl$ value of 62.8‰. A general ascending trend of isotope fractionation extents was found from the relatively lower-chlorinated PCDDs (Penta-CDD) to the higher-chlorinated (OCDD), with the range of $\Delta'^{37}Cl$ values from 20.8‰ to 62.8‰. The isotope fractionation extents of Hexa-CDFs, Hepta-CDFs and OCDF were similar to some extent, with the $\Delta'^{37}Cl$ values within 37.5‰-48.5‰. HCB,

Me-TCS, o,p'-DDE, and o,p'-DDD have the similar isotope fractionation extents with the $\Delta'^{37}Cl$ values from 43.9‰ to 55.0‰. It is notable that the isotope fractionation extents of o,p'-DDE, and o,p'-DDD were significant higher than those of their respective isomers p,p'-DDE and p,p'-DDD, with the discrepancies of 22.2‰ and 33.2‰, respectively. However, o,p'-DDT and p,p'-DDT had very similar isotope fractionation extents, with the $\Delta'^{37}Cl$ values of 37.4‰ and 36.0‰, respectively.

The extents of on-column bromine isotope fractionation of the investigated brominated compounds (except BDE-77) were generally lower than those of most of the chlorinated compounds exhibiting inverse isotope fractionation (Figure 3, G1-G3). As documented in Table S-4, the $\Delta'^{81}Br$ values of $^{13}C_6$-HBB, HBB and OBDD were similar and within the range of 15.5‰-19.7‰, and that of BDE-77 were relatively higher (38.7‰).

**Tentative Mechanistic Interpretation.**

*Conventional Explanations for On-column Isotope Fractionation.* According to the theories of Born-Oppenheimer approximation and simple harmonic oscillator model, the intramolecular bonds involving heavier isotopes have lower vibration frequencies, higher bond energies and slightly shorter lengths compared to those with lighter ones. The slightly shorter bonds result in the smaller molecular volumes of the heavier isotopomers than the lighter ones. The smaller molecular volumes reduce the dipole-induced polarisability of the heavier isotopomers to the stationary phase of GC columns. And the intermolecular interaction between the heavier isotopomers and the phenyl groups (relatively more polar groups of the stationary phase compared with the siloxane groups) is thus weakened. As a result, the heavier isotopomers elute faster than the lighter ones.

Normal chlorine isotope fractionation of p,p'-DDT on a pcGC column was reported previously.[36] The material of stationary phase of the pcGC column used in that study was the same as that used in our study (5% phenyl polysilphenylene-siloxane). The authors considered the normal chlorine isotope fractionation might be attributable to two reasons. The first reason was that the molecular volume differences between the heavier chlorine isotopologues and the lighter were probably smaller compared with those of carbon or hydrogen isotopologues; and the second was that the dominant motion process of p,p'-DDT on the pcGC column was transport in the mobile phase (carrier gas) under the conditions applied in that study (fast temperature program and megabore column), instead of the transfer process between the mobile and stationary phases.

However, in this study, we found that most (25 species) of the investigated compounds (including p,p'-DDT) presented inverse isotope fractionation on the analytical GC columns. Furthermore, the compounds presenting normal isotope fractionation and those exhibiting none isotope fractionation on the GC columns were found. Therefore, different HOPs could exhibit varied isotope fractionation behaviors under the same chromatographic conditions. Accordingly, this compound-specific isotope fractionation could not be simply explained with the above-mentioned two reasons proposed in the literature.[36]

*CG-model and Modified Two-Film Model.* Julien et al. applied the Craig-Gordon model (CG-model) in association with a two-film model to interpreting the isotope effect during the evaporation of 10 organic liquids under four evaporation modes.[11] This well-accepted model demonstrates that two isotope effects, i.e., liquid-vapor isotope effects ($\triangledown_{liq\text{-}vap}$) and diffusive isotope effects ($\varepsilon_{diff\text{-}He}$), act on the water evaporation process and are combinable. Thus the overall isotope effect can be expressed as:

$$IE = \nabla_{liq-vap} + \varepsilon_{diff-He} \qquad (4)$$

The two-film model, a conceptual model best suitable to represent the volatilization of liquids to the open air, hypothesizes two stagnant films, i.e., a liquid film on the liquid side of the interface and a gas film on the air side.[11] In accordance with the physical conditions, any one or both of the two films could contribute to the rate-limiting steps of volatilization. In this study, the stationary phase coated on the inner wall of the GC columns is a liquid film, which can be regarded as the liquid compartment in the two-film model. And the mobile phase (He) can be assumed as the air compartment. The two-film model for liquid evaporation is slightly different from the transfer process of compounds on GC columns. The evaporated gaseous molecules are belong to the same compound as the liquid in the two-film model, while the transferred molecules on the GC columns are belong to the investigated compound, and the liquid film is the mixture of the stationary phase (solvent) and the investigated compound (solute). The acting forces and modes are similar in the two-film model of liquid volatilization and in the transfer process of compound between stationary phase and carrier gas. Accordingly, we proposed a modified two-film model to tentatively elucidate the mechanisms of the chlorine/bromine isotope fractionation behaviors of HOPs on GC columns (Figure 5).

As shown in this modified model, the total isotope effects ($\varepsilon_{total}$) were composite effects of the $\triangledown_{liq\text{-}vap}$ isotope effects and $\varepsilon_{diff\text{-}He}$ isotope effects. Liquid-vapor isotope effects have been well studied, and most organic liquids exhibit inverse liquid-vapor isotope effects in terms of carbon isotopologues.[11] These isotope effects are dependent on the intermolecular interactions (van der Waals forces) including dipole-dipole force, induction force and dispersion (London) force. The heavier isotopologues have more compact molecular volumes (van der Waals volumes), which result in slightly weaker natural dipole moments and induced dipole moments. As a result, the dipole-dipole force, induction force and dispersion force, which depend on the natural dipole moments and/or induced dipole moments, become slightly smaller between the heavier isotopologues and the phenyl groups of column stationary phase. Therefore, the stationary phase of the GC columns possesses relatively lower adsorption effect or dissolving capacity for the heavier isotopologues than for the lighter ones. The heavier isotopologues thus are more liable to escape (volatilize) from the stationary phase into carrier gas than the lighter ones. Vise verse, the heavier isotopologues are more difficult to dissolve into the stationary phase from carrier gas than the lighter ones. The stationary phase containing the investigated compounds can be regarded as a solution system, of which the solutes are the investigated compounds and the solvent is the stationary phase (Figure 5). With a large number ($10^3$-$10^6$) of volatilizing-dissolving cycles (referring to theoretical plates), the heavier isotopologues thus could run faster than the lighter ones on the GC columns. As a consequence, the inverse isotope effects could take place.

On the other hand, the diffusion effect in the film on carrier gas side also plays a role in the isotope fractionation process on GC columns. The diffusion process is mass-dependent and determined by the differences among the molecular weights of isotopologues. The fractionation factor ($\alpha_{diff\text{-}He}$) derived from the diffusion effect in the film on carrier gas side can be calculated with Equation 5 (modified on the basis of the reported equation in the literature[37]):

$$\alpha_{diff-He} = \sqrt{\frac{M_l(M_h + M_{He})}{M_h(M_l + M_{He})}} \qquad (5)$$

where $M_l$ and $M_h$ are the molecular weights of a lighter isotopologue and a heavier isotopologue of an investigated compound, respectively and $M_{He}$ is the molecular weight of helium. And the diffusion isotope effects can be calculated with Equation 6 (modified on the basis of the reported equation in the literature[38]):

$$\varepsilon_{diff-He} = n(1-h)(\alpha_{diff-He} - 1)\cdot 1000\ ‰ \qquad (6)$$

where $n$ is a factor correcting for carrier gas flow (ranging from 1 to 0.5), and h is corresponding to the relative vapor saturation of organic compounds. As can be seen from Equation 6, the diffusion isotope effects ($\varepsilon_{diff\text{-}He}$) are always negative values, due to that the $\alpha_{diff\text{-}He}$ values are less than 1. This indicates the diffusion isotope effects are always normal. In the gas film, the intermolecular distances between the gaseous molecules are far larger than those between the liquid molecules in the liquid film. With the large intermolecular distances, the intermolecular interaction is very weak and negligible. It has been reported that the lighter isotopologues always have higher vapor pressure and higher diffusion rate in comparison with the heavier ones.[36,39,40] Therefore, in the He-diffusive sub-layer (Figure 5), the lighter isotopologues diffuse faster than the heavier ones, and thus are liable to enrich in the carrier gas film. As a result, normal diffusive isotope effects occur. When entering the layer of turbulently mixed carrier gas and the investigated compounds, the lighter isotopologues and the heavier reach isotope equilibrium.

As Figure 5 shows, the total isotope effects are the composition of the inverse $\bigtriangledown_{liq\text{-}vap}$ isotope effects and the normal $\varepsilon_{diff\text{-}He}$ isotope effects. When the composite isotope effects are inverse (green curves and arrows in Figure 5), the inverse isotope fractionation is present. The on-column inverse isotope fractionation of 29 HOPs observed in this study might follow this mode. If the composite isotope effects are normal (red curves and arrows in Figure 5), then normal isotope fractionation occurs. In this study, the observed normal isotope fractionation of PCB-138 and PCB-152 likely belonged to this case. If the absolute values of $\bigtriangledown_{liq\text{-}vap}$ isotope effects and $\varepsilon_{diff\text{-}He}$ isotope effects are equal (purple curves and arrows in Figure 5), then no isotope fractionation takes place. This mode can be applied to interpreting the finding that no observable isotope fractionation on GC columns was found for the four HOPs (i.e., PCB-101, PCB-180, TCDF and Penta-CDF-1) in this study.

**Implications for CSIA-Cl/Br Study.** Precision and accuracy are critical requirements for good CSIA methods which are of increasing interest in environmental sciences. The results obtained in this study indicated that most of the investigated HOPs presented significant chlorine/bromine isotope fractionation on GC columns. Thus, the chromatographic peak shape and the peak area integration may impact the precision and accuracy of CSIA-Cl/Br methods. The chromatographic peak in good symmetry and with suitable width is ideal. Asymmetric and large-width chromatographic peak could lead to difficulty for precise peak area integration, thus resulting in imprecise CSIA results. It is notable that the chromatographic peak with too small width could result in imprecise results for CSIA methods using GC–qMS, GC–DFS-MS and GC–QTOF-MS, due to the short dwell time for individual ion and/or insufficient acquisition points for a whole peak. Chromatographic peaks (except inseparable peaks) should be integrated as complete as possible in order to enhance the precision and accuracy of CSIA methods. Partial integration of chromatographic peaks will impair the precision and accuracy of CSIA not only for the methods using GC–qMS, GC–DFS-MS and GC–QTOF-MS (single-collector MSs), but also for those using GC-IRMS (multi-collector MS).

The observation in this study reveals that the relative isotope ratio variations ($\Delta^{h}E$) referenced to the overall isotope ratio of a compound were large in the both ends of the chromatographic peak but insignificant in the middle of the peak. In other words, the isotope ratio of an HOP derived from the middle retention-time segment is most close to the overall isotope ratio. If some compounds are not sufficiently separated, then it would be reasonable to integrate the middle retention-time segments of chromatographic peaks for obtaining relatively more accurate and precise isotope ratios.

In conclusion, CSIA-Cl/Br methods are still in their infancy in environmental applications involving degradation pathway elucidation and source appointment for HOPs. Therefore, it is important to evaluate the instrumental performances and set appropriate analytical schemes to develop precise, accurate, practical, convenient and cost-effective CSIA-Cl/Br methods for routine applications in future environmental studies. This study investigated the chlorine and bromine isotope fractionation on GC columns of 35 HOPs including 31 organochlorine compounds and four organobromine compounds. Different compounds exhibited varied isotope fractionation behaviors. Most of the HOPs (29 species) presented inverse isotope fractionation, and only two compounds (i.e., PCB-138 and PCB-153) exhibited normal isotope fractionation. The rest four compounds (i.e., PCB-101, Penta-CDF-1, PCB-180 and TCDF) presented no isotope fractionation. The isotope fractionation extents were varied among different HOPs, with the highest $\Delta'^{37}Cl$ value of 73.1‰ (PCB-18). The isotope fractionation extents of DDEs, DDDs, Tri-PCBs, and Penta-CDFs were significantly isomer-different. The mechanisms of the chlorine and bromine isotope fractionation on GC columns were tentatively elucidated with the CG-model and a modified two-film model. Two types of isotope effects, inverse $\nabla_{liq\text{-}vap}$ isotope effects and normal $\varepsilon_{diff\text{-}He}$ isotope effects, might do contribution to the total isotope effects together. The vector magnitudes of the combination of the two direction-opposite isotope effects determined the directions and extents of isotope fractionation. Thus, inverse, normal and unobservable isotope fractionation could present on GC columns for the HOPs. The suggestions based on the findings of this study for the future CSIA-Cl/Br research was proposed. The chromatographic peak with satisfactory symmetry and suitable width could be helpful to obtain precise and accurate data using CSIA-Cl/Br methods. The chromatographic peak area should be integrated as complete as possible to reduce the deviations of CSIA results, except for unseparated peaks. With respect to unseparated chromatographic peaks, calculating the isotope ratios with the middle retention-time segments would be conducive to achieve more reasonable CSIA-Cl/Br results.

## ASSOCIATED CONTENT

The Supporting Information is available free of charge on the ACS Publications website at http://pending.

## ACKNOWLEDGEMENTS

We are grateful for our colleagues Dr. Lianjun Bao and Dr. Man Ren, for their kind gifts of some reference standards. This work was partially supported by the National Natural Science Foundation of China (Grant No. 41603092).

**Legends**

**Figures**

**Figure 1.** Schematic workflow of the evaluation methods for revealing chlorine and bromine isotope fractionation of HOPs on GC columns. TIC: total ion chromatogram.

**Figure 2.** Isotope ratios of the investigated HOPs derived from different retention-time segments (T1-T5). G1-G7: compound groups divided based on the isotope ratios as well as isotope fractionation extents and directions.

**Figure 3.** Relative isotope ratio variations ($\Delta^{h}E$) referenced to overall isotope ratios of the HOPs derived from different retention-time segments (T1-T5). G1-G6: compound groups divided based on the isotope fractionation extents and directions.

**Figure 4.** Chlorine and bromine isotope fractionation extents ($\Delta'^{h}E$) of the HOPs on GC columns. Orange bars: compounds exhibiting significant inverse isotope fractionation ($\Delta'^{h}E$ > 10.0‰); Green bars: compounds presenting unobservable isotope fractionation (-5.0‰ < $\Delta'^{h}E$ < 5.0‰); Yellow bar: compound exhibiting low normal isotope fractionation (-10.0‰ < $\Delta'^{h}E$ < -5.0‰); Blue bar: compound exhibiting significant normal isotope fractionation ($\Delta'^{h}E$ < -10.0‰).

**Figure 5.** Schematic illustration of the modified two-film model for isotope effects of HOPs undergoing volatilizing-dissolving separation cycles on GC columns with rate limitation on the boundary of carrier gas (He) side.

**Figures**

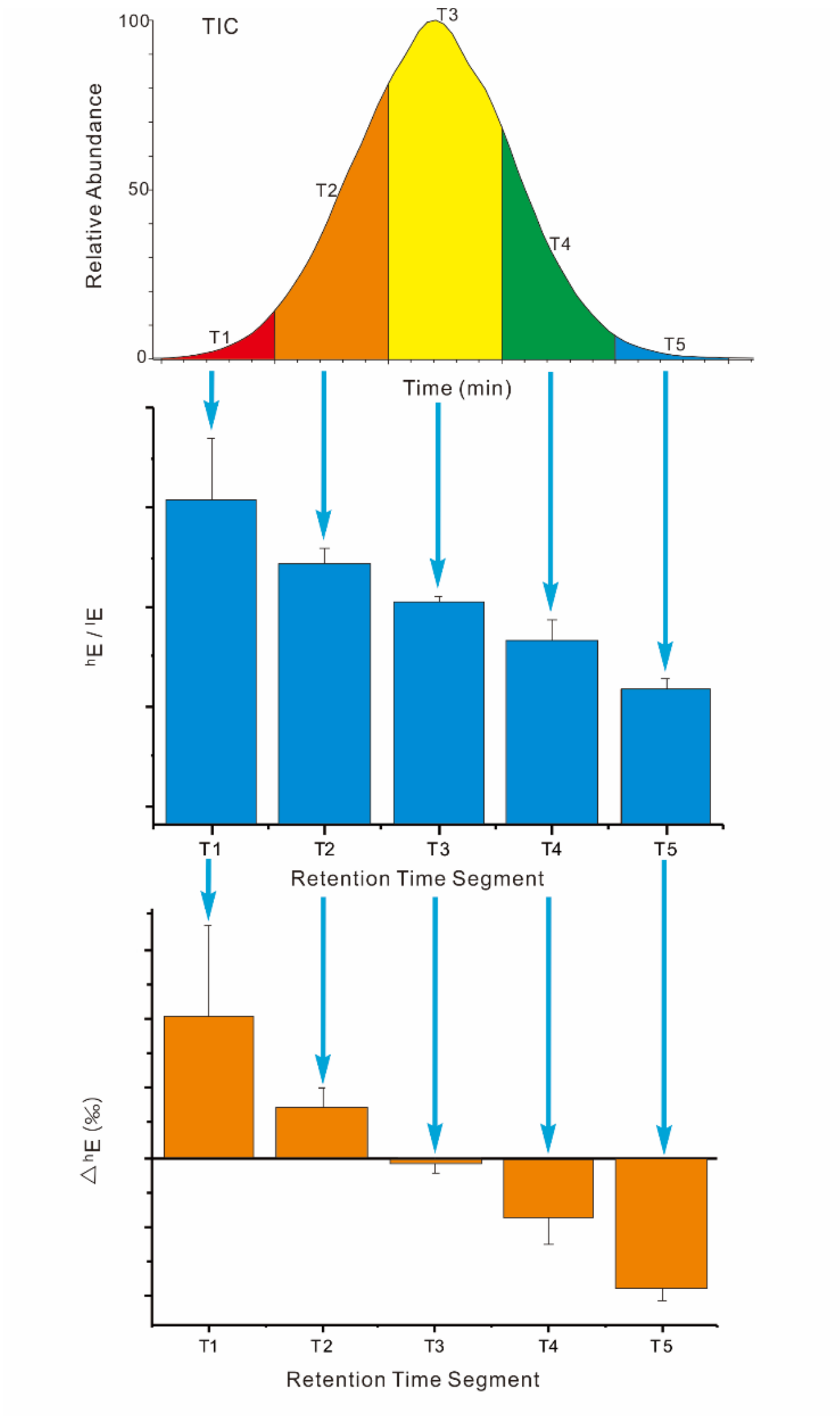

**Figure 1.** Schematic workflow of the evaluation methods for revealing chlorine and bromine isotope fractionation (isotope fractionation) of HOPs on GC columns. TIC: total ion chromatogram.

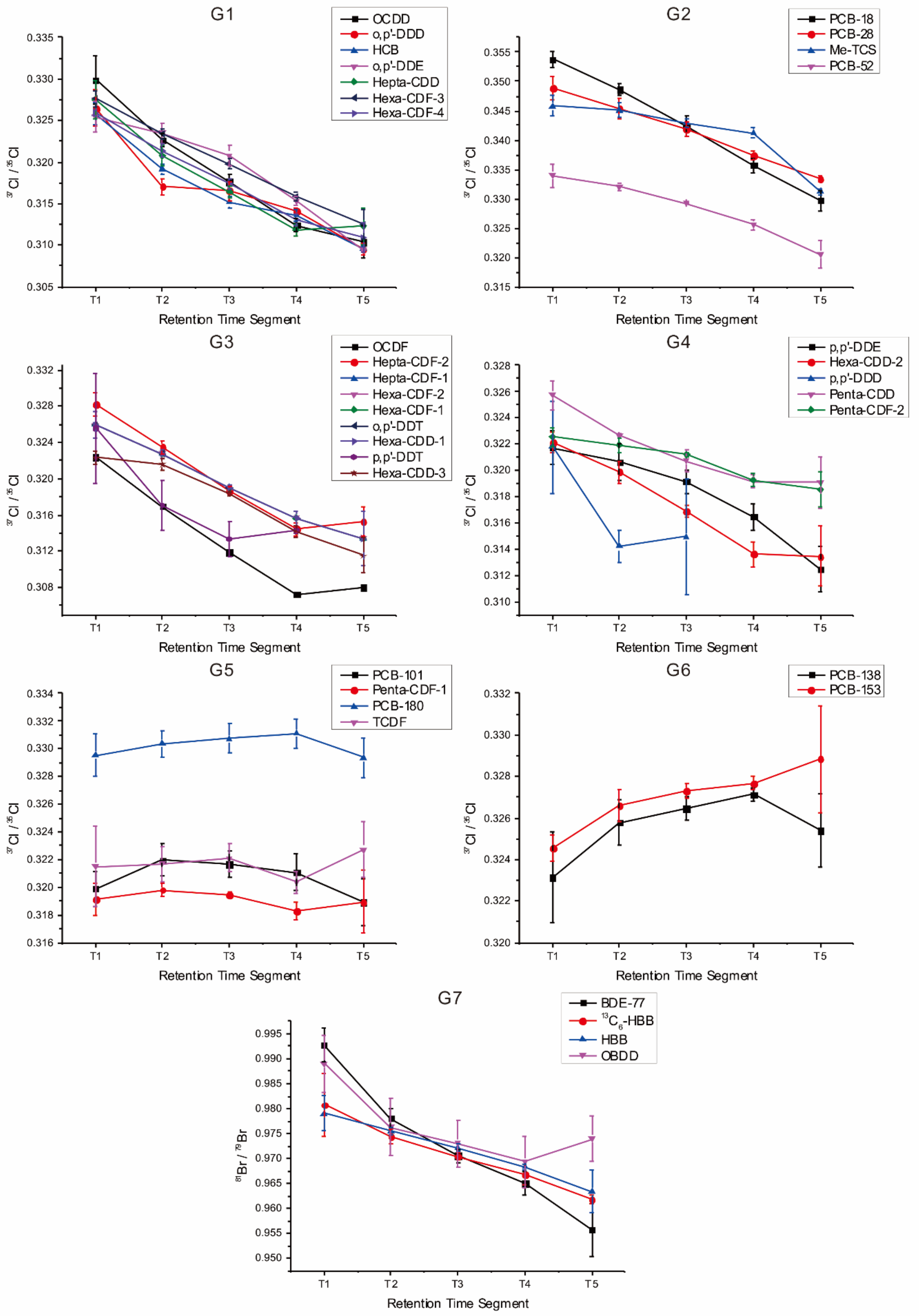

**Figure 2.** Isotope ratios of the investigated HOPs derived from different retention-time segments (T1-T5). G1-G7: compound groups divided based on the isotope ratios as well as isotope fractionation extents and directions.

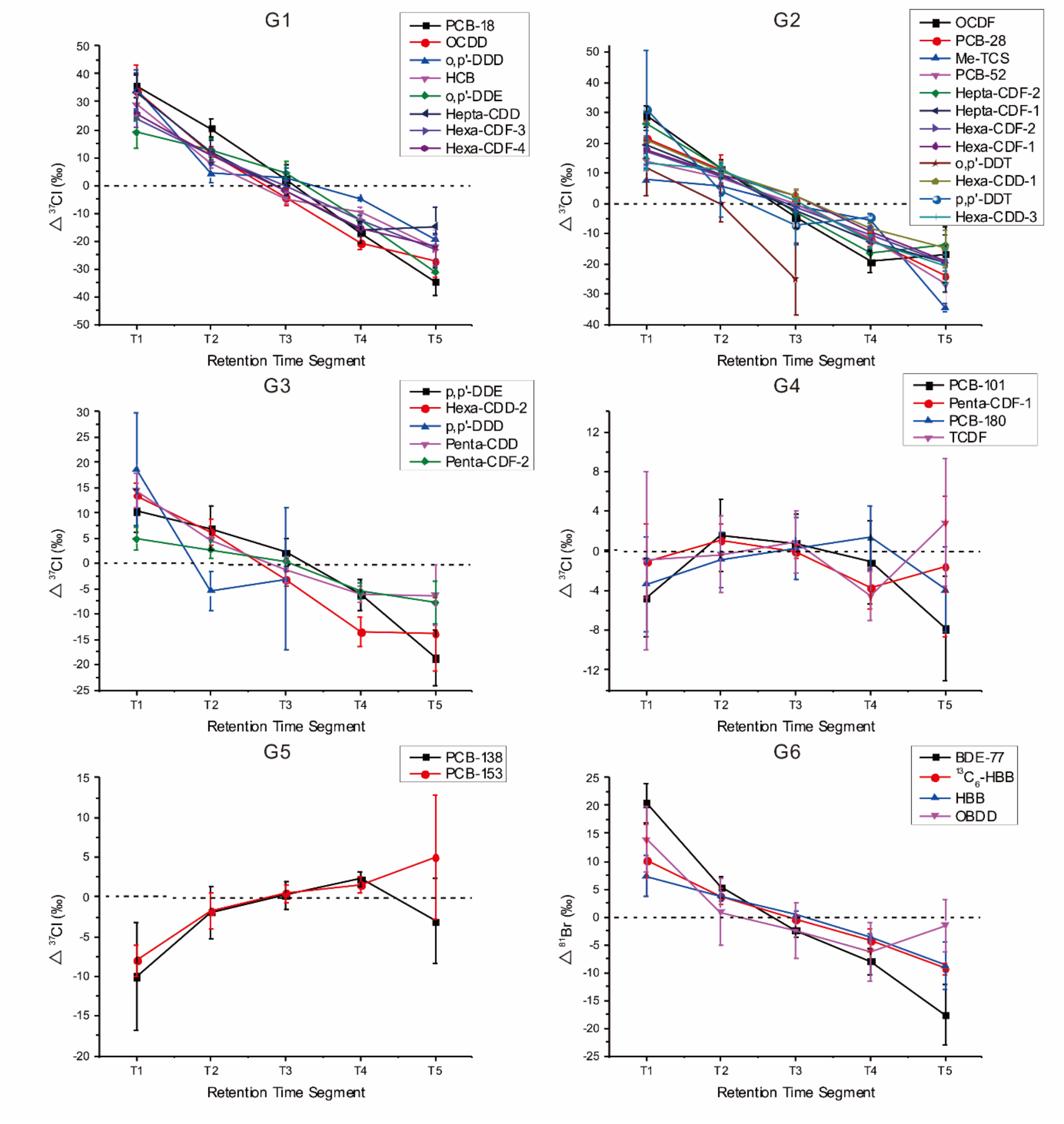

**Figure 3.** Relative isotope ratio variations ($\Delta^{h}E$) referenced to overall isotope ratios of the HOPs derived from different retention-time segments (T1-T5). G1-G6: compound groups divided based on the isotope fractionation extents and directions.

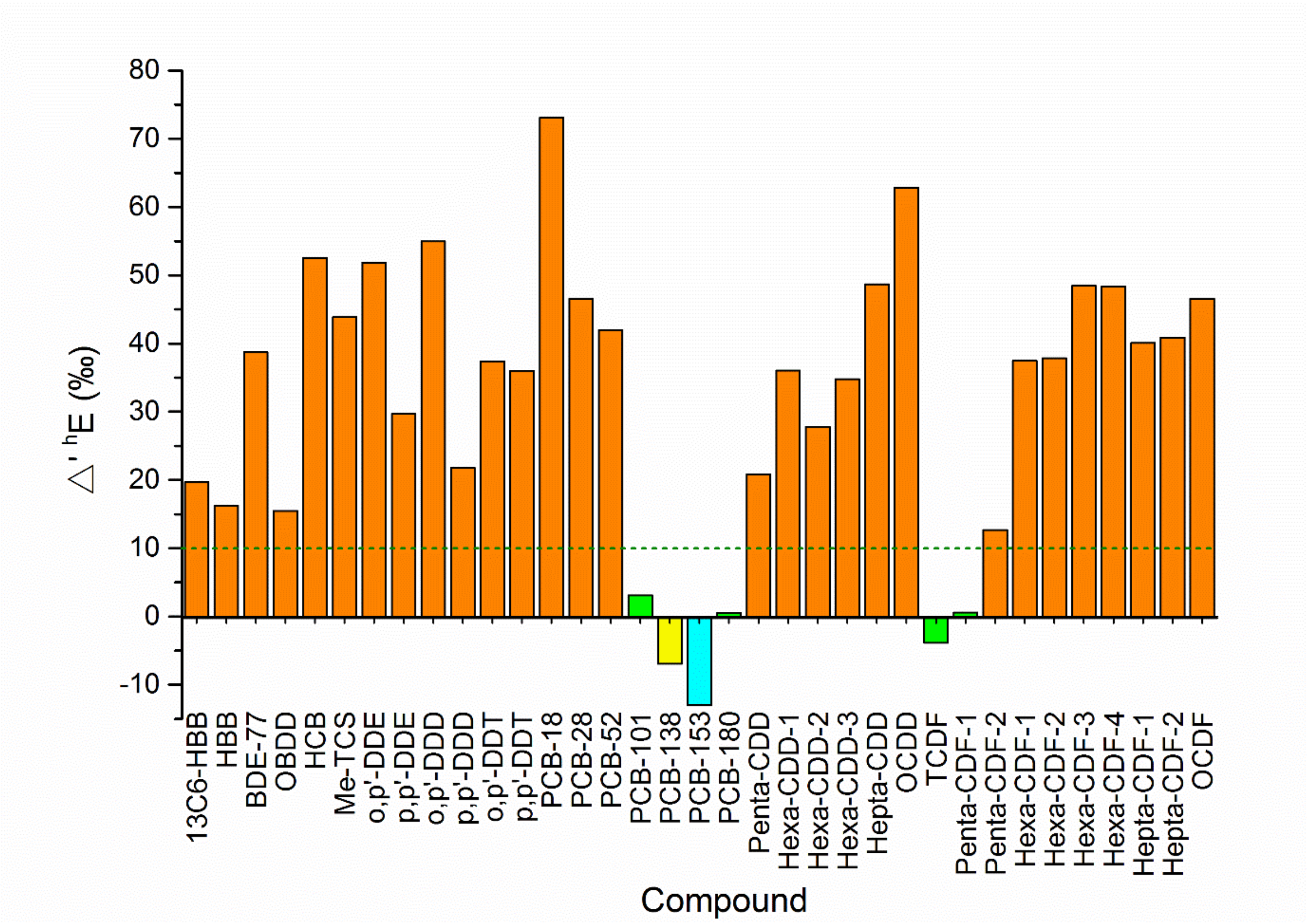

**Figure 4.** Chlorine and bromine isotope fractionation extents ($\Delta'^{h}E$) of the HOPs on GC columns. Orange bars: compounds exhibiting significant inverse isotope fractionation ($\Delta'^{h}E$ > 10.0‰); Green bars: compounds presenting unobservable isotope fractionation (-5.0‰ < $\Delta'^{h}E$ < 5.0‰); Yellow bar: compound exhibiting low normal isotope fractionation (-10.0‰ < $\Delta'^{h}E$ < -5.0‰); Blue bar: compound exhibiting significant normal isotope fractionation ($\Delta'^{h}E$ < -10.0‰).

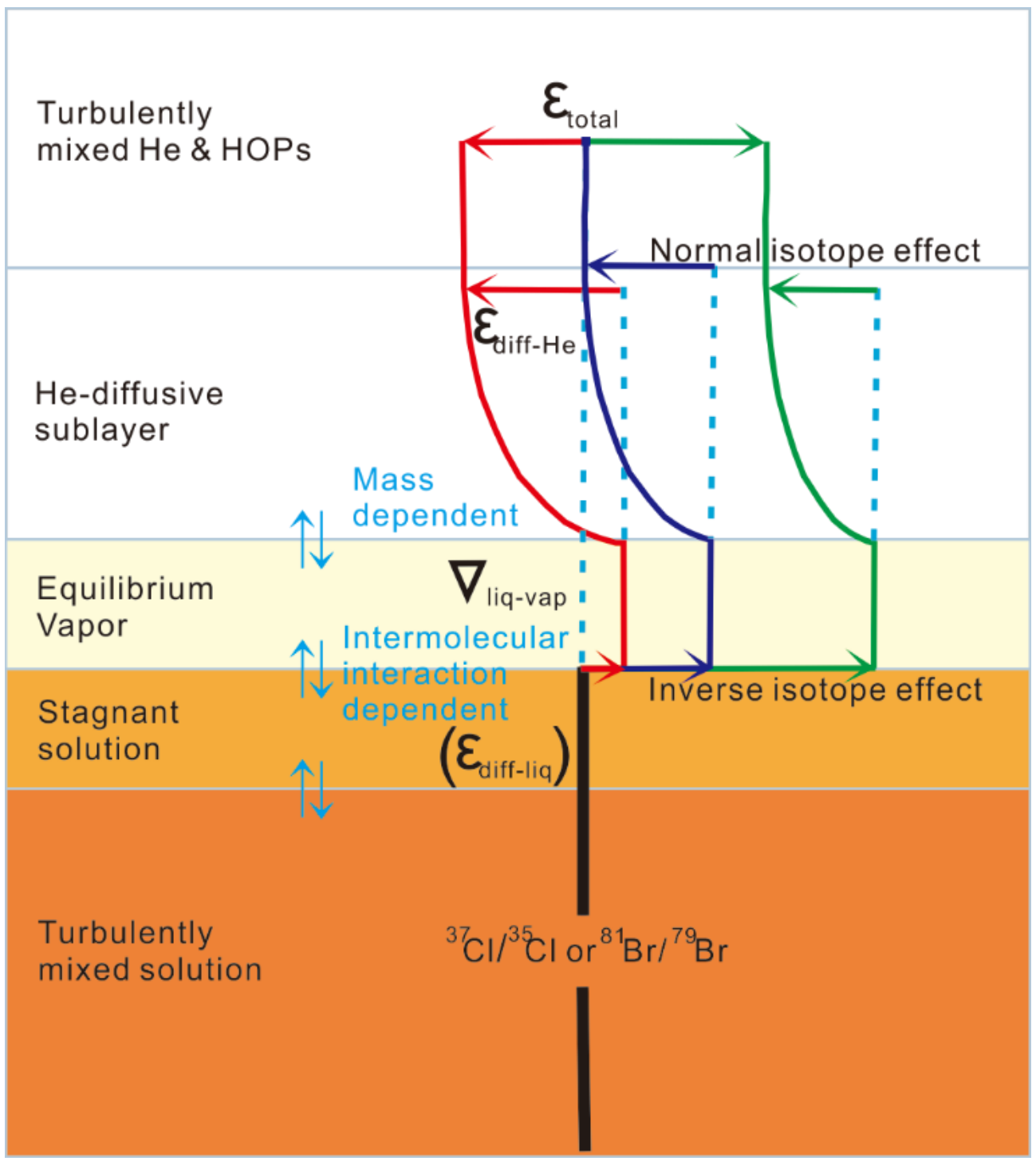

**Figure 5.** Schematic illustration of the modified two-film model for isotope effects of HOPs undergoing volatilizing-dissolving separation cycles on GC columns with rate limitation on the boundary of carrier gas (He) side.

TOC Graphic

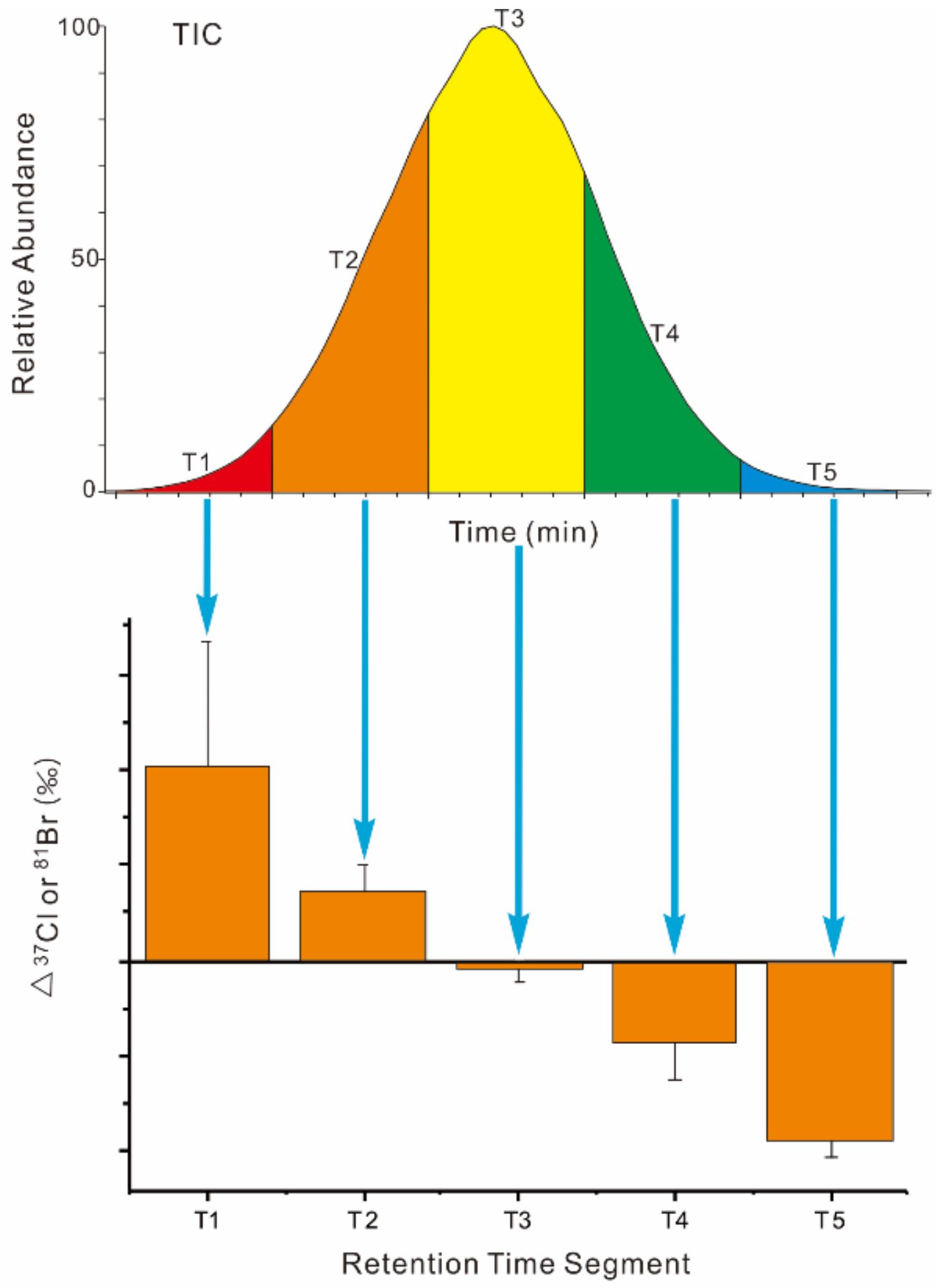

# SUPPORTING INFORMATION

## Chlorine and Bromine Isotope Fractionation of Halogenated Organic Pollutants on Gas Chromatography Columns

**Caiming Tang[1,2], Jianhua Tan[3], Songsong Xiong[1,2], Jun Liu[1,2], Yujuan Fan[1,2], Xianzhi Peng[1,*]**

[1] *State Key Laboratory of Organic Geochemistry, Guangzhou Institute of Geochemistry, Chinese Academy of Sciences, Guangzhou 510640, China*

[2] *University of Chinese Academy of Sciences, Beijing 100049, China*

[3] *Guangzhou Quality Supervision and Testing Institute, Guangzhou, 510110, China*

*Corresponding Author.

Tel: +86-020-85290009; Fax: +86-020-85290009. E-mail: pengx@gig.ac.cn. (X. Peng).

## ASSOCIATED CONTENT

Considering the space limit, we present some additional information and tables as noted in the text in the *Supporting Information*. This material is available via the Internet at http://pending.

## ADDITIONAL EXPERIMENTAL SECTION

**Chemicals and Materials.** Reference standards including $^{13}C_6$-hexabromobenzene ($^{13}C_6$-HBB), octabromodibenzo-p-dioxin (OBDD), calibration standard solution of chlorinated dioxins/furans (CS5, containing 17 types of PCDD/Fs), perfluorotributylamine (FC43) and perfluorokerosene (PFK) were bought from Cambridge Isotope Laboratories Inc. (Andover, MA, USA). Standard solutions of 3,3',4,4'-tetrabrominated biphenyl ether (BDE-77), polychlorinated biphenyls (PCB-18, PCB-28, PCB-52, PCB-101, PCB-138, PCB-153 and PCB-180), pesticides (o,p'-DDT, p,p'-DDT, o,p'-DDD, p,p'-DDD, o,p'-DDE, and p,p'-DDE) were purchased from AccuStandard Inc. (New Haven, CT, USA). Methyl-triclosan (99.5%, Me-TCS) and hexachlorobenzene (99.5%, HCB) were bought from Dr. Ehrenstorfer (Augsburg, Germany). Full names, abbreviations, CAS No. and structures of the chemicals are listed in Table S-1.

HPLC-grade solvents including nonane and isooctane were bought from Alfa Aesar Company (Ward Hill, MA, USA) and CNW Technologies GmbH (Düsseldorf, Germany), respectively.

**Stock and Working Solutions.** All the purchased standards (except Me-TCS and HCB) were in the form of either mixed or individual solution prepared with solvents such as nonane, toluene and isooctane. Pure standards Me-TCS (liquid) and HCB (powder) were accurately weighed and dissolved in isooctane to prepare stock solutions with the concentration of 1.0 mg/mL. Except the PCDD/Fs calibration solution (CS5), all the rest purchased standard solutions and prepared stock solutions were further diluted with either nonane or isooctane to get working solutions with appropriate concentrations suitable for GC-HRMS analysis (Table S-1). All standard solutions were kept in a freezer at -20 $^{o}$C before use.

**Additional Data Processing.** The isotope ratios were also reported as differences in "per mil" (‰), in the "delta notation" ($\delta^{h}$E) referenced to the Standard Mean Ocean Element (SMOE):

$$\delta^{h}\mathrm{E} = \left( \frac{R_{sample}}{R_{SMOE}} - 1 \right) \cdot 1000‰ \qquad (2)$$

where E represents elements Cl or Br; $^{h}$E and $^{l}$E represent the heavy isotope ($^{37}$Cl or $^{81}$Br) and the light ($^{35}$Cl or $^{79}$Br), respectively; R is the ratio of $^{h}$E/$^{l}$E ($^{37}$Cl/$^{35}$Cl or $^{81}$Br/$^{79}$Br).

## ADDITIONAL RESULTS AND DISCUSSION

**Isotope Ratio Analysis Method.** CSIA-Cl and CSIA-Br are usually conducted with GC separation followed by off-line or on-line (continuous flow)-IRMS detection.[1,2,3] Recently, some CSIA-Cl methods have been developed using commonly used GC–qMS and GC–QTOF-MS for organochlorine pollutants such as trichloroethene (TCE), perchloroethene (PCE), and HCB.[4,5] In the present study, we applied GC–DFS-HRMS to the method development of compound-specific chlorine/bromine isotope analysis (CSIA-Cl/Br) for HOPs. The advantages of DFS-HRMS including high resolution and sensitivity could provide high selectivity and signal intensity for these CSIA-Cl/Br methods. Unlike IRMS, however, DFS-HRMS cannot acquire isotope ratio data directly, so do qMS and QTOF-MS. Therefore, mathematical data analysis is required to obtain isotope ratios of chlorine and bromine from mass spectra generated by DFS-HRMS, qMS and QTOF-MS. To date, several evaluation schemes for calculating chlorine isotope ratios based on mass spectra derived from qMS have been reported, including molecular ion method, conventional multiple ion method, modified multiple ion method, and complete ion method.[4] On the basis of these previously reported evaluation schemes and in light of the performance features of DFS-MS, we developed a modified calculation method, i.e., complete molecular-ion method (Equation 1). DFS-MS in MID mode is very suitable for monitoring molecular ions, due to its high sensitivity and selectivity. On the other hand, DFS-HRMS is unsuitable to simultaneously detect multiple ions covering a relatively large mass range in one MID segment, for instance, molecular ions and their product ions. Thus, the multiple ion methods and complete ion method, which were reported to be more precise than the molecular ion method in CSIA-Cl, could be inappropriate for CSIA-Cl/Br study using DFS-HRMS.

With the developed GC–DFS-HRMS detection method and the complete molecular-ion scheme of isotope ratio calculation, satisfactory results (SD ≤ 0.5‰) of CSIA-Cl/Br for a majority of the investigated HOPs (22 out of 35) were achieved (Table S-4). The SDs (n=5) of the isotope ratios of 22 compounds were ≤ 0.5‰, and those of 7 compounds were within the range of 0.51‰-0.97‰. The isotope ratio SDs of the rest 6 compounds were in the range of 1.08‰-2.22‰. Therefore, the precision of our CSIA method could fulfil the requirement for investigating Cl/Br isotope fractionation of HOPs on GC columns.

**Tables**

**Table S-1.** Names, chemical information, concentrations and chromatographic separation conditions of the investigated compounds.

**Table S-2.** Retention times, chemical formulas, isotopologue formulas, exact molecular weights and exact *m/z* values of the investigated compounds.

**Table S-3.** Isotope ratios, delta values ($\delta^hE$, referenced to SMOE) and relative variations ($\Delta^hE$) derived from different retention time segments of the investigated compounds. SMOE: Standard Mean Ocean Element.

**Table S-4.** Overall isotope ratios and $\delta^hE$ values (referenced to SMOE) and isotope fractionation extents ($\Delta'^hE$) of all the investigated compounds along with precision results of the developed CSIA method.

## Tables

**Table S-1.** Names, chemical information, concentrations and chromatographic separation conditions of the investigated compounds.

| Compound | Abbreviation | Structure | CAS No. | Column | Temperature program | Concentration (ng/mL) | Injection solvent |
|---|---|---|---|---|---|---|---|
| $^{13}C_6$-Hexabromobenzene | $^{13}C_6$-HBB | 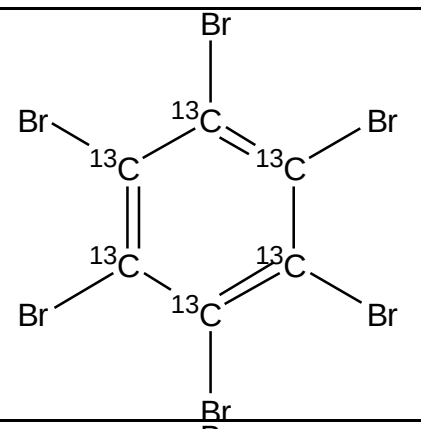 | 85380-74-1 | Long[a] | A program (Inlet: 260 $^{o}$C; Transfer line: 280 $^{o}$C) | 1000 | Isooctane |
| Hexabromobenzene | HBB | 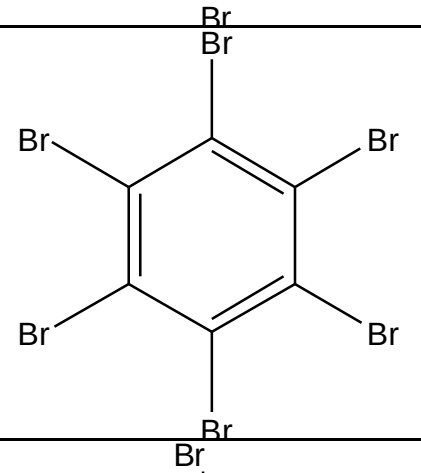 | 87-82-1 | Long | A program (Inlet: 260 $^{o}$C; Transfer line: 280 $^{o}$C) | NA (Soil sample) | Nonane |
| 3,3',4,4'-Tetrabrominated biphenyl ether | BDE-77 | 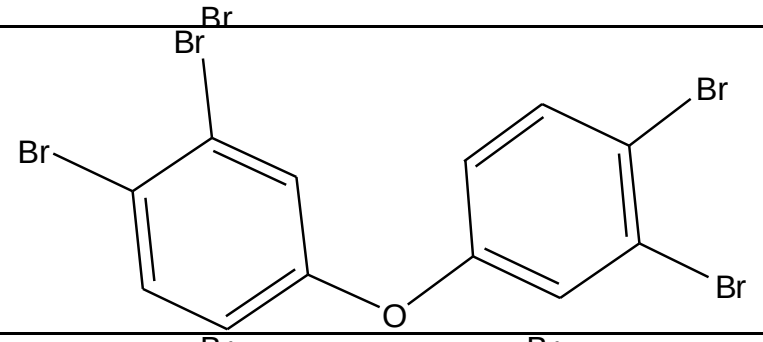 | 93703-48-1 | Long | B program (Inlet: 260 $^{o}$C; Transfer line: 280 $^{o}$C) | 500 | Nonane |
| 1,2,3,4,6,7,8,9-Octabromodibenzo-p-dioxin | OBDD | 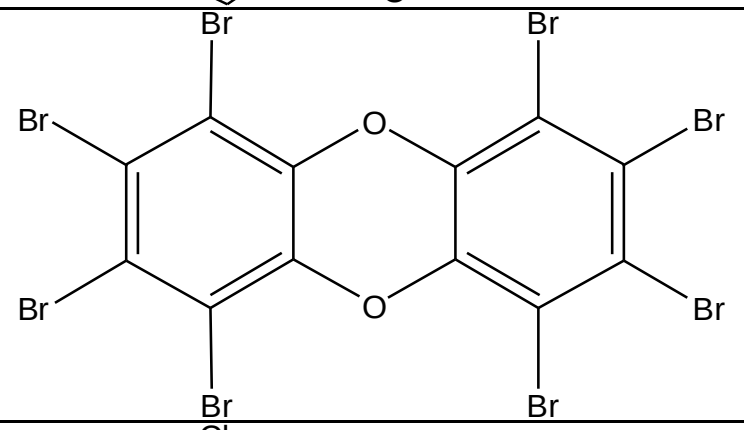 | 2170-45-8 | Short[b] | C program (Inlet: 280 $^{o}$C; Transfer line: 300 $^{o}$C) | 5000 | Toluene |
| Hexachlorobenzene | HCB | 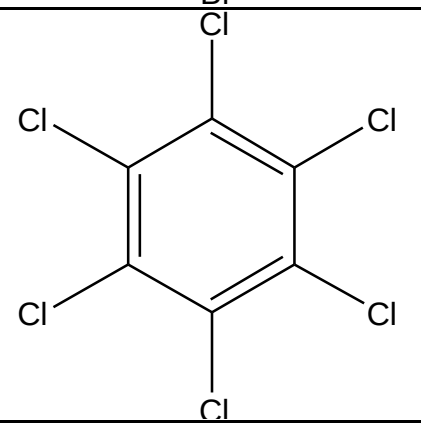 | 118-74-1 | Long | A program (Inlet: 260 $^{o}$C; Transfer line: 280 $^{o}$C) | 446 | Isooctane |

| Compound | Abbreviation | Structure | CAS No. | Column | Temperature program | Concentration (ng/mL) | Injection solvent |
|---|---|---|---|---|---|---|---|
| Methyl-triclosan | Me-TCS | 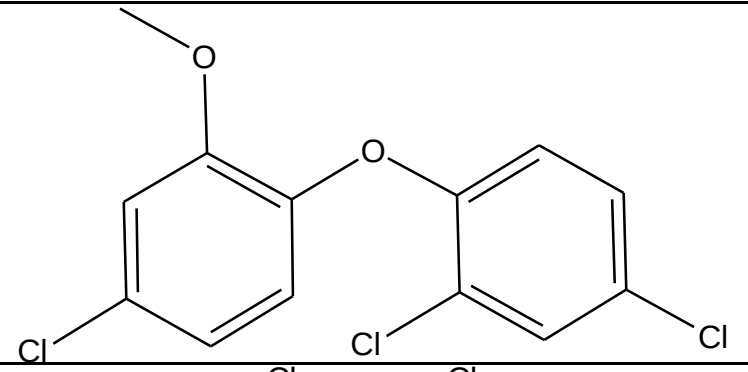 | 4640-01-1 | Long | D program (Inlet: 260 $^{o}$C; Transfer line: 280 $^{o}$C) | 1000 | Nonane |
| o,p'-Dichlorodiphenyldichloroethylene | o,p'-DDE | 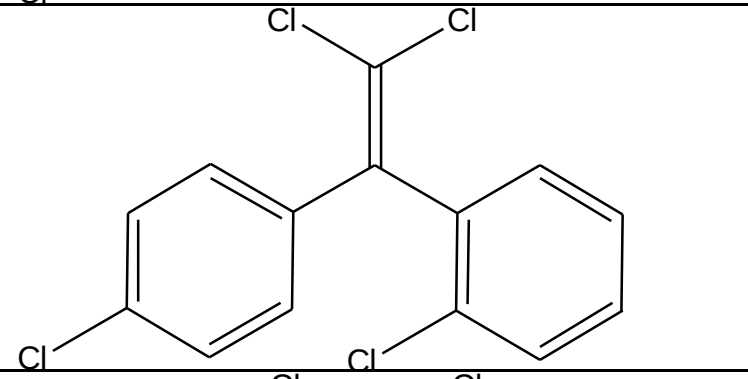 | 3424-82-6 | Long | D program (Inlet: 260 $^{o}$C; Transfer line: 280 $^{o}$C) | 1000 | Nonane |
| p,p'-Dichlorodiphenyldichloroethylene | p,p'-DDE | 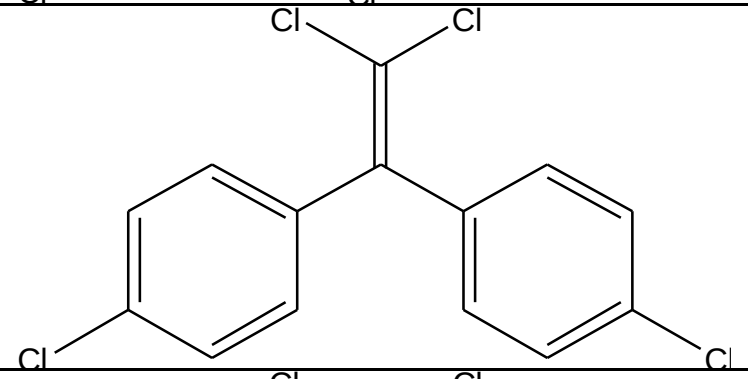 | 72-55-9 | Long | D program (Inlet: 260 $^{o}$C; Transfer line: 280 $^{o}$C) | 1000 | Nonane |
| 2,2-Bis(2-chlorophenyl-4-chlorophenyl)-1,1-dichloroethane | o,p'-DDD | 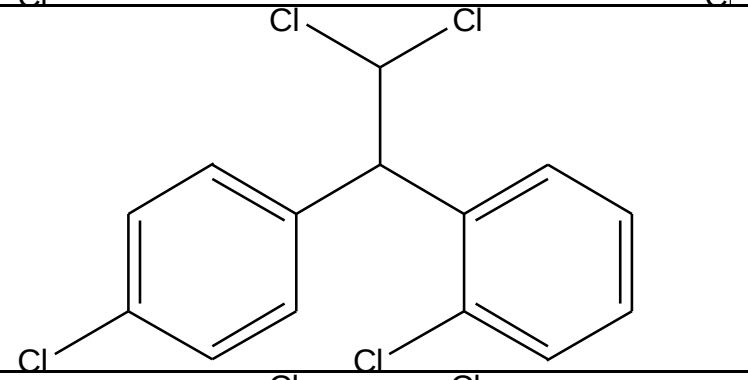 | 53-19-0 | Long | D program (Inlet: 260 $^{o}$C; Transfer line: 280 $^{o}$C) | 5000 | Nonane |
| 2,2-Bis(4-chlorophenyl-4-chlorophenyl)-1,1-dichloroethane | p,p'-DDD | 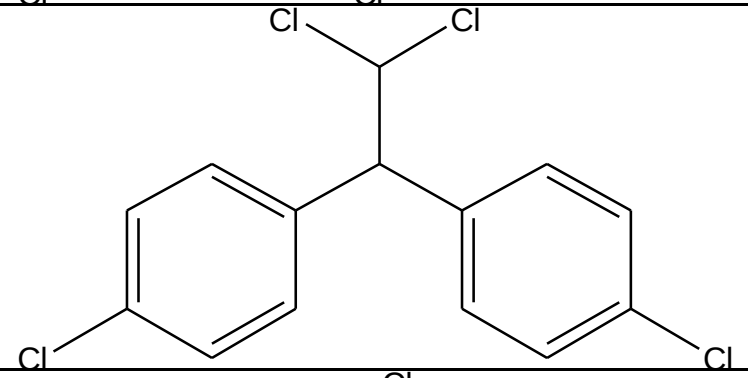 | 72-54-8 | Long | D program (Inlet: 260 $^{o}$C; Transfer line: 280 $^{o}$C) | 5000 | Nonane |
| o,p'-Dichlorodiphenyltrichloroethane | o,p'-DDT | 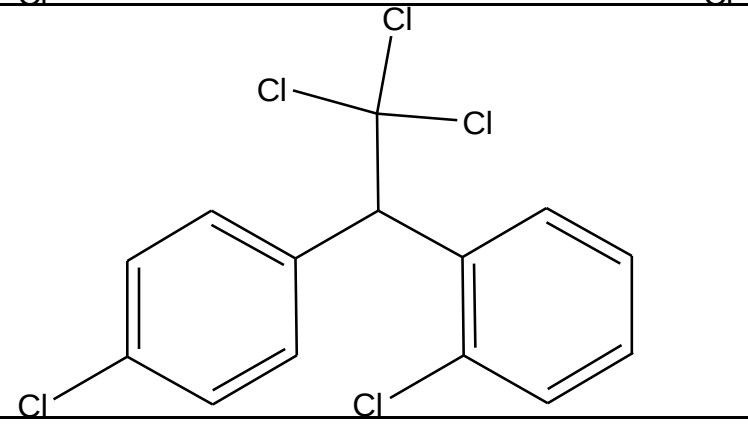 | 789-02-6 | Long | D program (Inlet: 260 $^{o}$C; Transfer line: 280 $^{o}$C) | 5000 | Nonane |

| Compound | Abbreviation | Structure | CAS No. | Column | Temperature program | Concentration (ng/mL) | Injection solvent |
|---|---|---|---|---|---|---|---|
| p,p'-Dichlorodiphenyltrichloroethane | p,p'-DDT | 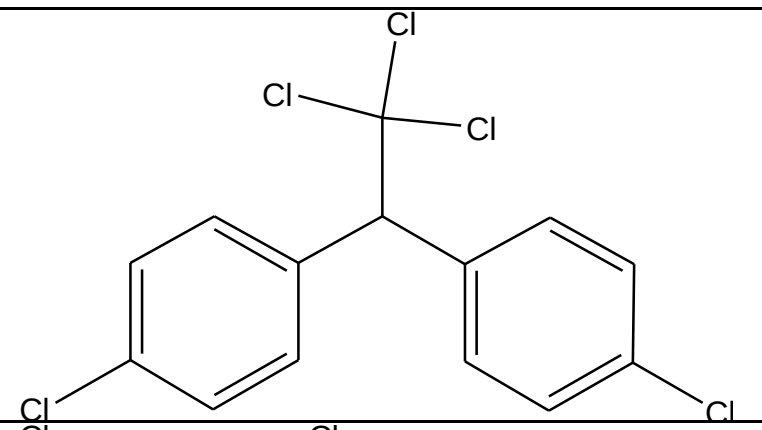 | 50-29-3 | Long | D program (Inlet: 260 ºC; Transfer line: 280 ºC) | 5000 | Nonane |
| 2,2'5-Trichloro-1,1'biphenyl | PCB-18 | 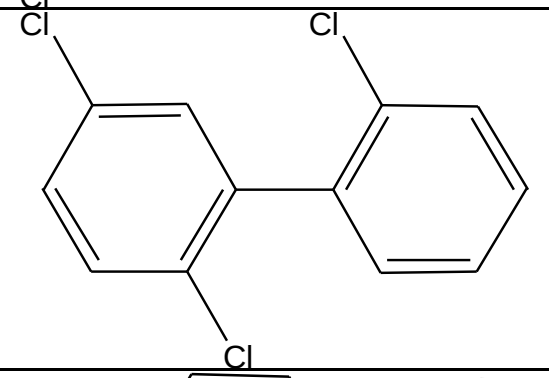 | 37680-65-2 | Long | A program (Inlet: 260 ºC; Transfer line: 280 ºC) | 1000 | Isooctane |
| 2,4,4'-Trichloro-1,1'biphenyl | PCB-28 | 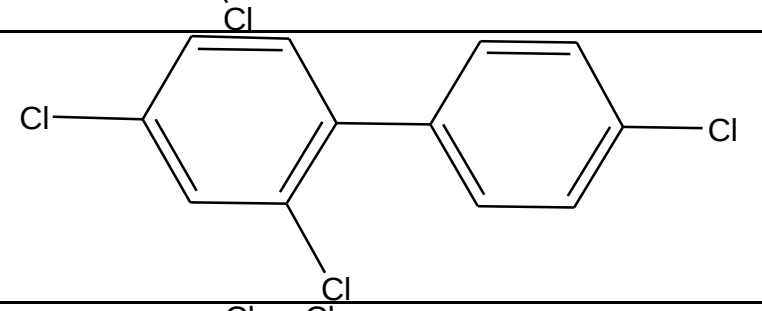 | 7012-37-5 | Long | A program (Inlet: 260 ºC; Transfer line: 280 ºC) | 1000 | Isooctane |
| 2,2',5,5'-Tetrachloro-1,1'biphenyl | PCB-52 | 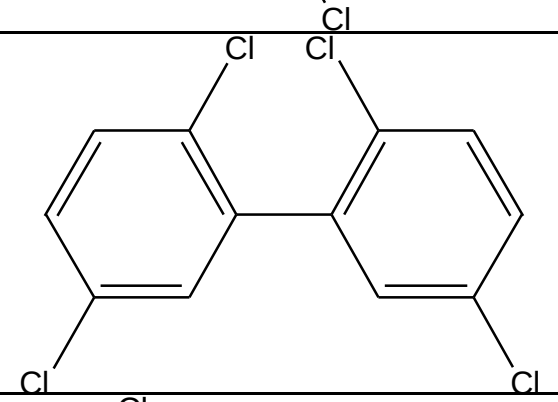 | 35693-99-3 | Long | A program (Inlet: 260 ºC; Transfer line: 280 ºC) | 1000 | Isooctane |
| 2,2',4,5,5'-Hentachloro-1,1'biphenyl | PCB-101 | 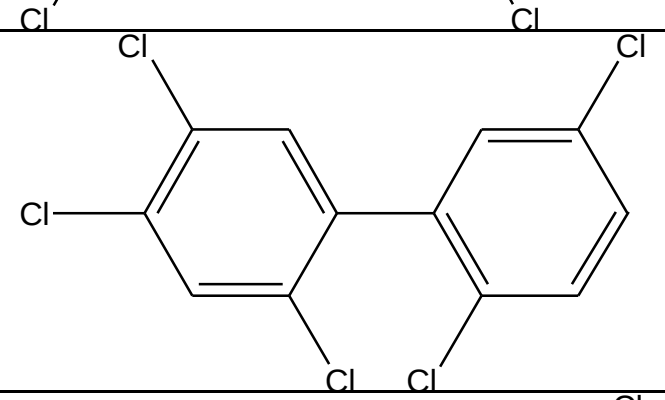 | 37680-73-2 | Long | A program (Inlet: 260 ºC; Transfer line: 280 ºC) | 1000 | Isooctane |
| 2,2',3,4,4',5'-Hexachloro-1,1'biphenyl | PCB-138 | 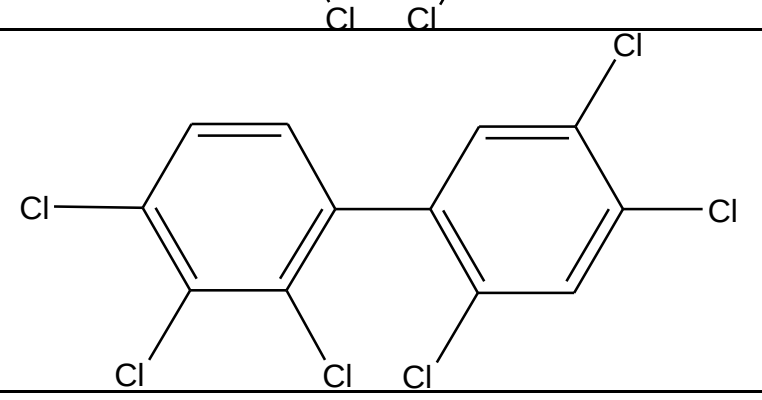 | 35065-28-2 | Long | A program (Inlet: 260 ºC; Transfer line: 280 ºC) | 1000 | Isooctane |

| Compound | Abbreviation | Structure | CAS No. | Column | Temperature program | Concentration (ng/mL) | Injection solvent |
|---|---|---|---|---|---|---|---|
| 2,2',4,4',5,5'-Hexachloro-1,1'biphenyl | PCB-153 | 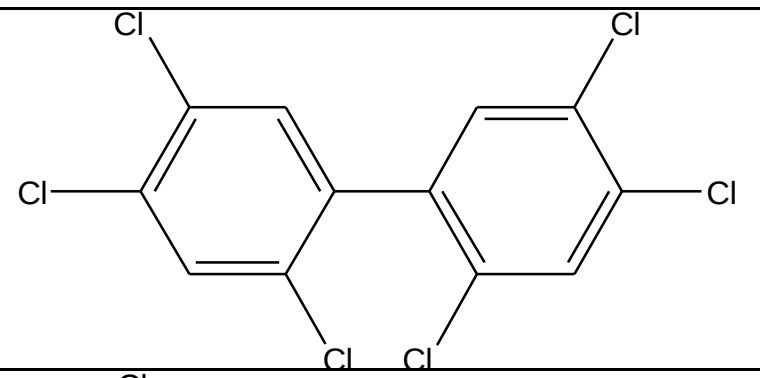 | 35065-27-1 | Long | A program (Inlet: 260 ºC; Transfer line: 280 ºC) | 1000 | Isooctane |
| 2,2',3,4,4',5,5'-Heptaachloro-1,1'biphenyl | PCB-180 | 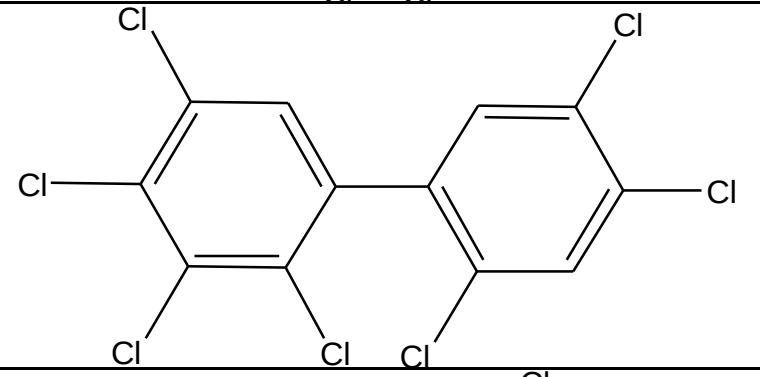 | 35065-29-3 | Long | A program (Inlet: 260 ºC; Transfer line: 280 ºC) | 1000 | Isooctane |
| 1,2,3,7,8-Pentachlorodibenzo-p-dioxin | Penta-CDD | 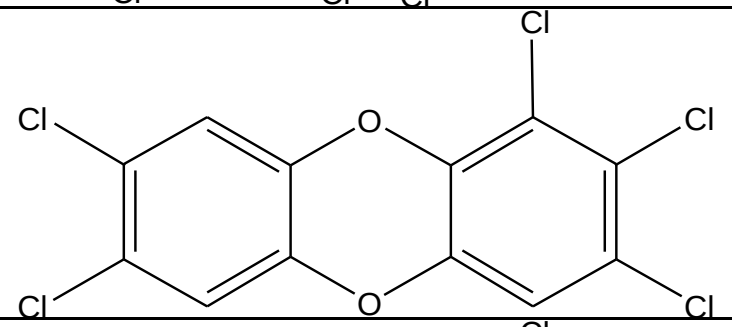 | 40321-76-4 | Long | E program (Inlet: 250 ºC; Transfer line: 280 ºC) | 1000 | Nonane |
| 1,2,3,4,7,8-Hexachlorodibenzo-p-dioxin | Hexa-CDD-1 | 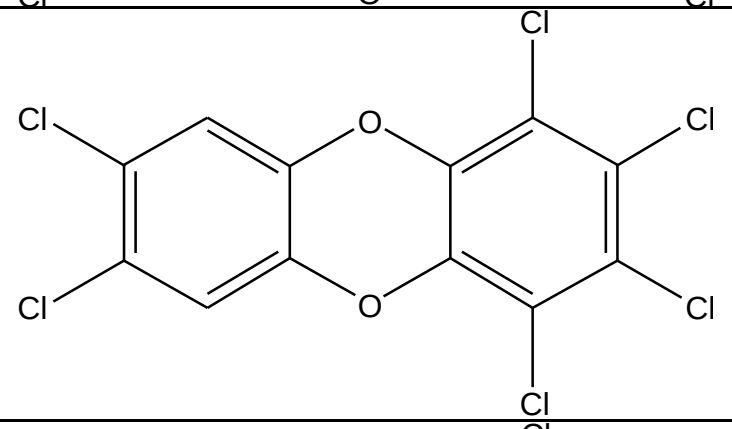 | 39227-28-6 | Long | E program (Inlet: 250 ºC; Transfer line: 280 ºC) | 1000 | Nonane |
| 1,2,3,6,7,8-Hexachlorodibenzo-p-dioxin | Hexa-CDD-2 | 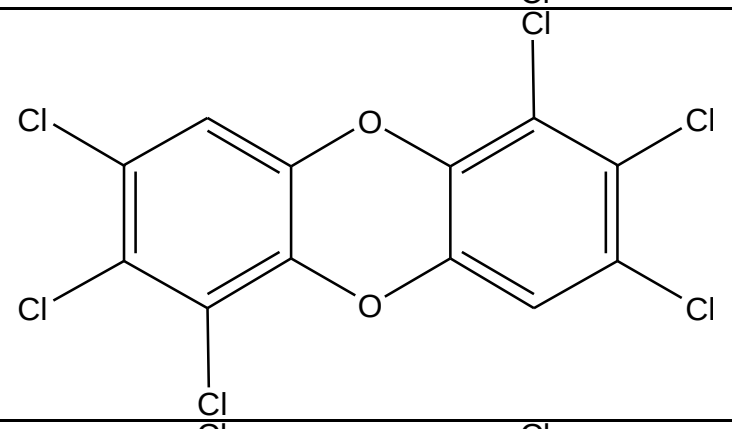 | 57653-85-7 | Long | E program (Inlet: 250 ºC; Transfer line: 280 ºC) | 1000 | Nonane |
| 1,2,3,7,8,9-Hexachlorodibenzo-p-dioxin | Hexa-CDD-3 | 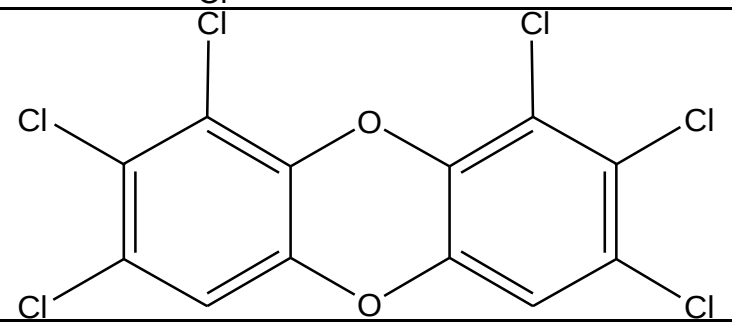 | 19408-74-3 | Long | E program (Inlet: 250 ºC; Transfer line: 280 ºC) | 1000 | Nonane |

| Compound | Abbreviation | Structure | CAS No. | Column | Temperature program | Concentration (ng/mL) | Injection solvent |
|---|---|---|---|---|---|---|---|
| 1,2,3,4,6,7,8-Heptachlorodibenzo-p-dioxin | Hepta-CDD | 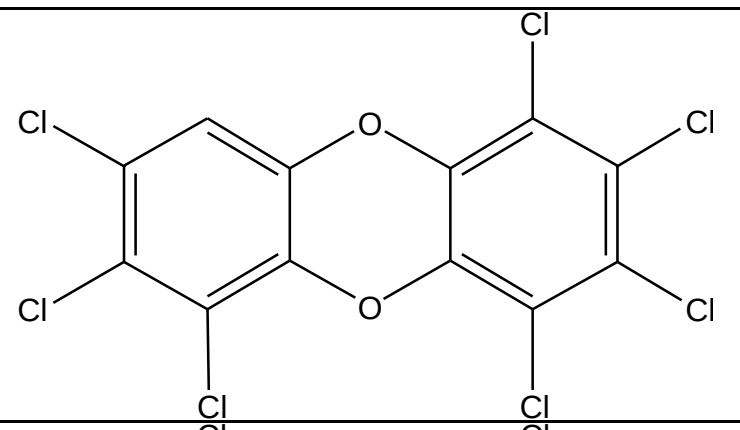 | 35822-46-9 | Long | E program (Inlet: 250 ºC; Transfer line: 280 ºC) | 1000 | Nonane |
| Octachlorodibenzo-p-dioxin | OCDD | 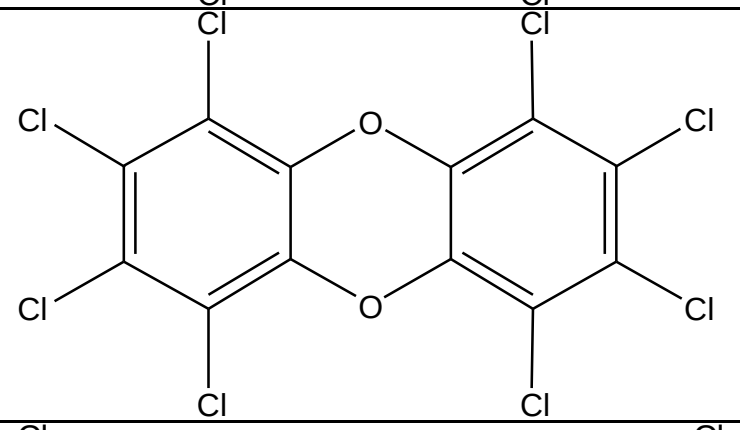 | 3268-87-9 | Long | E program (Inlet: 250 ºC; Transfer line: 280 ºC) | 2000 | Nonane |
| 2,3,7,8-Tetrachlorodibenzofuran | TCDF | 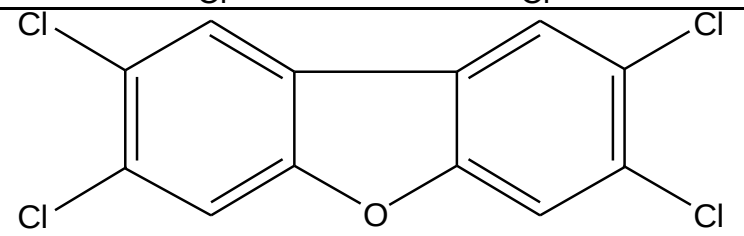 | 51207-31-9 | Long | E program (Inlet: 250 ºC; Transfer line: 280 ºC) | 200 | Nonane |
| 1,2,3,7,8-Pentachlorodibenzofuran | Penta-CDF-1 | 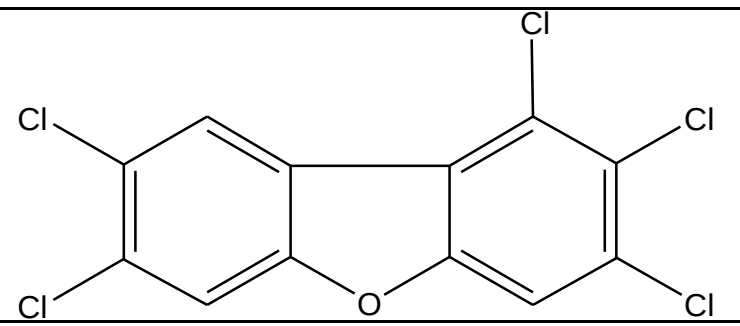 | 57117-41-6 | Long | E program (Inlet: 250 ºC; Transfer line: 280 ºC) | 1000 | Nonane |
| 2,3,4,7,8-Pentachlorodibenzofuran | Penta-CDF-2 | 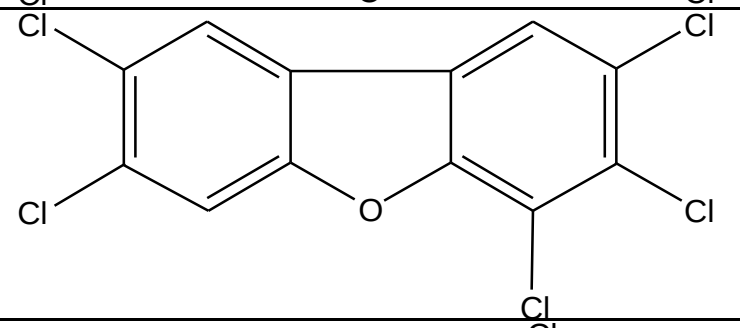 | 57117-31-4 | Long | E program (Inlet: 250 ºC; Transfer line: 280 ºC) | 1000 | Nonane |
| 1,2,3,4,7,8-Hexachlorodibenzofuran | Hexa-CDF-1 | 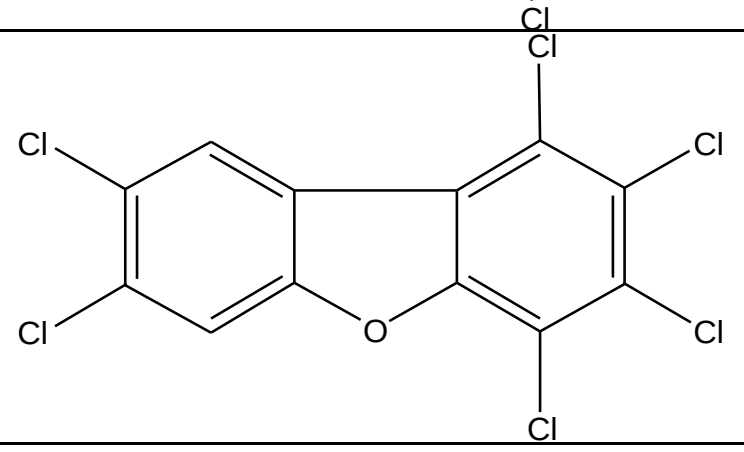 | 55684-94-1 | Long | E program (Inlet: 250 ºC; Transfer line: 280 ºC) | 1000 | Nonane |

| Compound | Abbreviation | Structure | CAS No. | Column | Temperature program | Concentration (ng/mL) | Injection solvent |
|---|---|---|---|---|---|---|---|
| 1,2,3,6,7,8-Hexachlorodibenzofuran | Hexa-CDF-2 | 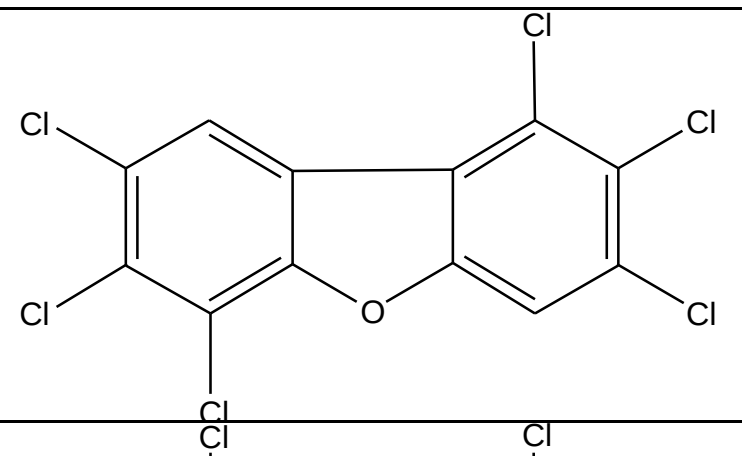 | 57117-44-9 | Long | E program (Inlet: 250 ºC; Transfer line: 280 ºC) | 1000 | Nonane |
| 1,2,3,7,8,9-Hexachlorodibenzofuran | Hexa-CDF-3 | 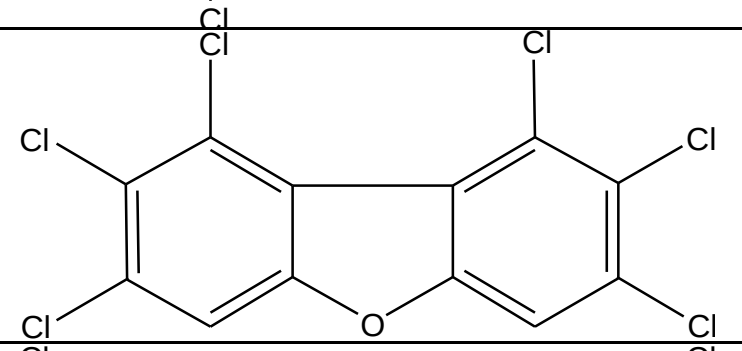 | 72918-21-9 | Long | E program (Inlet: 250 ºC; Transfer line: 280 ºC) | 1000 | Nonane |
| 2,3,4,6,7,8-Hexachlorodibenzofuran | Hexa-CDF-4 | 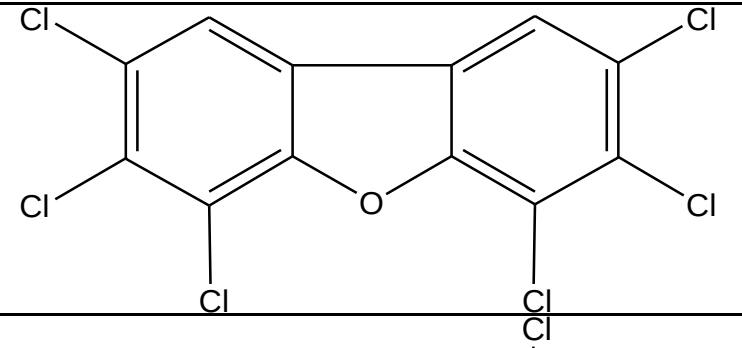 | 60851-34-5 | Long | E program (Inlet: 250 ºC; Transfer line: 280 ºC) | 1000 | Nonane |
| 1,2,3,4,6,7,8-Heptachlorodibenzofuran | Hepta-CDF-1 | 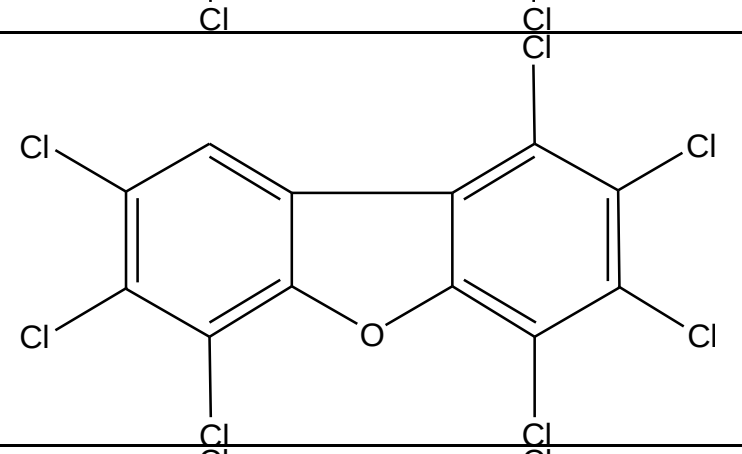 | 67652-39-5 | Long | E program (Inlet: 250 ºC; Transfer line: 280 ºC) | 1000 | Nonane |
| 1,2,3,4,7,8,9-Heptachlorodibenzofuran | Hepta-CDF-2 | 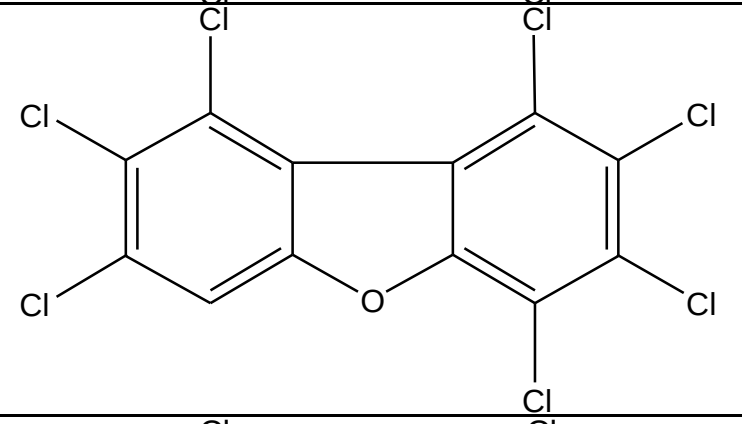 | 55673-89-7 | Long | E program (Inlet: 250 ºC; Transfer line: 280 ºC) | 1000 | Nonane |
| Octachlorodibenzofuran | OCDF | 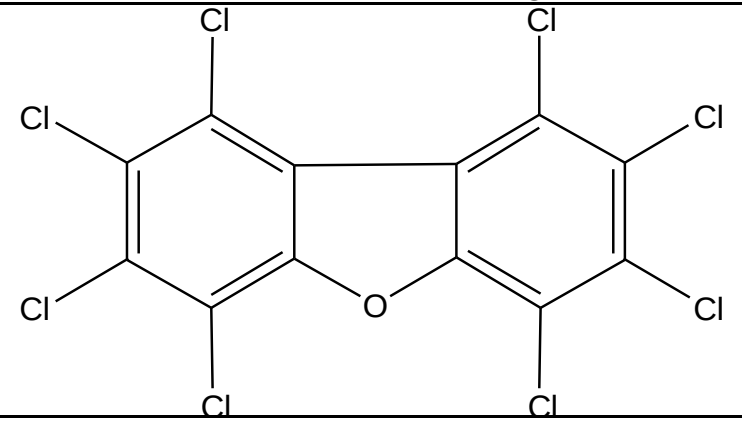 | 39001-02-0 | Long | E program (Inlet: 250 ºC; Transfer line: 280 ºC) | 2000 | Nonane |

Note, columns and temperature-raising programs:

a: 60 m × 0.25 mm, 0.25 μm thickness, J&W Scientific, USA; b: 30 m × 0.25 mm, 0.1 μm thickness, J&W Scientific, USA;
A: held at 120 ℃ for 2 min, ramped to 220 ℃ at 20 ℃/min, held for 16 min, then ramped to 235 ℃ at 5 ℃/min, held for 7 min, and then ramped to 260 ℃ at 5 ℃/min, and finally ramped to 330 ℃ at 30 ℃/min and held for 9.67 min;
B: held at 120 ℃ for 2 min, ramped to 220 ℃ at 20 ℃/min, held for 16 min, then ramped to 235 ℃ at 5 ℃/min, held for 7 min, and then ramped to 260 ℃ at 5 ℃/min, and finally ramped to 330 ℃ at 30 ℃/min and held for 19.67 min;
C: held at 120 ℃ for 1 min, ramped to 160 ℃ at 40 ℃/min, then ramped to 235 ℃ at 25 ℃/min, held for 2 min, and then ramped to 280 ℃ at 15 ℃/min, held for 10 min, and finally ramped to 315 ℃ at 10 ℃/min and held for 11.5 min;
D: held at 120 ℃ for 2 min, ramped to 220 ℃ at 20 ℃/min, held for 16 min, then ramped to 235 ℃ at 5 ℃/min, held for 7 min, and finally ramped to 330 ℃ at 30 ℃/min and held for 3.83 min;
E: held at 120 ℃ for 2 min, ramped to 220 ℃ at 20 ℃/min, held for 16 min, then ramped to 235 ℃ at 5 ℃/min, held for 7 min, and finally ramped to 330 ℃ at 5 ℃/min and held for 8 min.

**Table S-2.** Retention times, chemical formulas, isotopologue formulas, exact molecular weights and exact *m/z* values of the investigated compounds.

| Compound | Retention time (min) | Formula | Isotopologue formula | Exact molecular weight (u) | Exact *m/z* value (u) |
|---|---|---|---|---|---|
| $^{13}C_6$-HBB | 37.95 | $^{13}C_6Br_6$ | $^{13}C_6{}^{79}Br_6$ | 551.53015 | 551.52960 |
| | | | $^{13}C_6{}^{79}Br_5{}^{81}Br$ | 553.52810 | 553.52755 |
| | | | $^{13}C_6{}^{79}Br_4{}^{81}Br_2$ | 555.52605 | 555.52550 |
| | | | $^{13}C_6{}^{79}Br_3{}^{81}Br_3$ | 557.52401 | 557.52346 |
| | | | $^{13}C_6{}^{79}Br_2{}^{81}Br_4$ | 559.52196 | 559.52141 |
| | | | $^{13}C_6{}^{79}Br{}^{81}Br_5$ | 561.51991 | 561.51936 |
| | | | $^{13}C_6{}^{81}Br_6$ | 563.51786 | 563.51731 |
| | | | | | |
| HBB | 38.16 | $C_6Br_6$ | $C_6{}^{79}Br_6$ | 545.51002 | 545.50947 |
| | | | $C_6{}^{79}Br_5{}^{81}Br$ | 547.50797 | 547.50742 |
| | | | $C_6{}^{79}Br_4{}^{81}Br_2$ | 549.50592 | 549.50537 |
| | | | $C_6{}^{79}Br_3{}^{81}Br_3$ | 551.50387 | 551.50332 |
| | | | $C_6{}^{79}Br_2{}^{81}Br_4$ | 553.50183 | 553.50128 |
| | | | $C_6{}^{79}Br{}^{81}Br_5$ | 555.49978 | 555.49923 |
| | | | $C_6{}^{81}Br_6$ | 557.49773 | 557.49718 |
| | | | | | |
| BDE-77 | 41.79 | $C_{12}H_6Br_4O$ | $C_{12}H_6{}^{79}Br_4O$ | 481.71521 | 481.71466 |
| | | | $C_{12}H_6{}^{79}Br_3{}^{81}BrO$ | 483.71316 | 483.71261 |
| | | | $C_{12}H_6{}^{79}Br_2{}^{81}Br_2O$ | 485.71112 | 485.71057 |
| | | | $C_{12}H_6{}^{79}Br{}^{81}Br_3O$ | 487.70907 | 487.70852 |
| | | | $C_{12}H_6{}^{81}Br_4O$ | 489.70702 | 489.70647 |
| | | | | | |
| OBDD | 28.99 | $C_{12}Br_8O_2$ | $C_{12}{}^{79}Br_8O_2$ | 807.33652 | 807.33597 |
| | | | $C_{12}{}^{79}Br_7{}^{81}BrO_2$ | 809.33447 | 809.33392 |
| | | | $C_{12}{}^{79}Br_6{}^{81}Br_2O_2$ | 811.33242 | 811.33187 |
| | | | $C_{12}{}^{79}Br_5{}^{81}Br_3O_2$ | 813.33038 | 813.32983 |
| | | | $C_{12}{}^{79}Br_4{}^{81}Br_4O_2$ | 815.32833 | 815.32778 |
| | | | $C_{12}{}^{79}Br_3{}^{81}Br_5O_2$ | 817.32628 | 817.32573 |
| | | | $C_{12}{}^{79}Br_2{}^{81}Br_6O_2$ | 819.32424 | 819.32369 |
| | | | $C_{12}{}^{79}Br{}^{81}Br_7O_2$ | 821.32219 | 821.32164 |
| | | | $C_{12}{}^{81}Br_8O_2$ | 823.32014 | 823.31959 |
| | | | | | |
| HCB | 11.69 | $C_6Cl_6$ | $C_6{}^{35}Cl_6$ | 281.81311 | 281.81256 |
| | | | $C_6{}^{35}Cl_5{}^{37}Cl$ | 283.81016 | 283.80961 |
| | | | $C_6{}^{35}Cl_4{}^{37}Cl_2$ | 285.80721 | 285.80666 |
| | | | $C_6{}^{35}Cl_3{}^{37}Cl_3$ | 287.80426 | 287.80371 |
| | | | $C_6{}^{35}Cl_2{}^{37}Cl_4$ | 289.80132 | 289.80077 |
| | | | $C_6{}^{35}Cl{}^{37}Cl_5$ | 291.79837 | 291.79782 |
| | | | $C_6{}^{37}Cl_6$ | 293.79542 | 293.79487 |
| | | | | | |
| Me-TCS | 20.47 | $C_{13}H_9Cl_3O_2$ | $C_{13}H_9{}^{35}Cl_3O_2$ | 301.96681 | 301.96626 |

| Compound | Retention time (min) | Formula | Isotopologue formula | Exact molecular weight (u) | Exact *m/z* value (u) |
|---|---|---|---|---|---|
| | | | $C_{13}H_{9}{}^{35}Cl_{2}{}^{37}ClO_{2}$ | 303.96386 | 303.96331 |
| | | | $C_{13}H_{9}{}^{35}Cl^{37}Cl_{2}O_{2}$ | 305.96091 | 305.96036 |
| | | | $C_{13}H_{9}{}^{37}Cl_{3}O_{2}$ | 307.95796 | 307.95741 |
| | | | | | |
| o,p’-DDE | 20.27 | $C_{14}H_{8}Cl_{4}$ | $C_{14}H_{8}{}^{35}Cl_{4}$ | 315.93801 | 315.93746 |
| | | | $C_{14}H_{8}{}^{35}Cl_{3}{}^{37}Cl$ | 317.93506 | 317.93451 |
| | | | $C_{14}H_{8}{}^{35}Cl_{2}{}^{37}Cl_{2}$ | 319.93211 | 319.93156 |
| | | | $C_{14}H_{8}{}^{35}Cl^{37}Cl_{3}$ | 321.92916 | 321.92861 |
| | | | $C_{14}H_{8}{}^{37}Cl_{4}$ | 323.92621 | 323.92566 |
| | | | | | |
| p,p’-DDE | 22.97 | $C_{14}H_{8}Cl_{4}$ | Refer to o,p’-DDE | | |
| | | | | | |
| o,p’-DDD | 26.52 | $C_{14}H_{10}Cl_{4}$ | $C_{14}H_{10}{}^{35}Cl_{4}$ | 317.95366 | 317.95311 |
| | | | $C_{14}H_{10}{}^{35}Cl_{3}{}^{37}Cl$ | 319.95071 | 319.95016 |
| | | | $C_{14}H_{10}{}^{35}Cl_{2}{}^{37}Cl_{2}$ | 321.94776 | 321.94721 |
| | | | $C_{14}H_{10}{}^{35}Cl^{37}Cl_{3}$ | 323.94481 | 323.94426 |
| | | | $C_{14}H_{10}{}^{37}Cl_{4}$ | 325.94186 | 325.94131 |
| | | | | | |
| p,p’-DDD | 26.73 | $C_{14}H_{10}Cl_{4}$ | Refer to o,p’-DDD | | |
| | | | | | |
| o,p’-DDT | 26.72 | $C_{14}H_{9}Cl_{5}$ | $C_{14}H_{9}{}^{35}Cl_{5}$ | 351.91469 | 351.91414 |
| | | | $C_{14}H_{9}{}^{35}Cl_{4}{}^{37}Cl$ | 353.91174 | 353.91119 |
| | | | $C_{14}H_{9}{}^{35}Cl_{3}{}^{37}Cl_{2}$ | 355.90879 | 355.90824 |
| | | | $C_{14}H_{9}{}^{35}Cl_{2}{}^{37}Cl_{3}$ | 357.90584 | 357.90529 |
| | | | $C_{14}H_{9}{}^{35}Cl^{37}Cl_{4}$ | 359.90289 | 359.90234 |
| | | | $C_{14}H_{9}{}^{37}Cl_{5}$ | 361.89994 | 361.89939 |
| | | | | | |
| p,p’-DDT | 29.94 | $C_{14}H_{9}Cl_{5}$ | Refer to o,p’-DDT | | |
| | | | | | |
| PCB-18 | 12.42 | $C_{12}H_{7}Cl_{3}$ | $C_{12}H_{7}{}^{35}Cl_{3}$ | 255.96133 | 255.96078 |
| | | | $C_{12}H_{7}{}^{35}Cl_{2}{}^{37}Cl$ | 257.95838 | 257.95783 |
| | | | $C_{12}H_{7}{}^{35}Cl^{37}Cl_{2}$ | 259.95543 | 259.95488 |
| | | | $C_{12}H_{7}{}^{37}Cl_{3}$ | 261.95248 | 261.95193 |
| | | | | | |
| PCB-28 | 14.06 | $C_{12}H_{7}Cl_{3}$ | Refer to PCB-18 | | |
| | | | | | |
| PCB-52 | 15.42 | $C_{12}H_{6}Cl_{4}$ | $C_{12}H_{6}{}^{35}Cl_{4}$ | 289.92236 | 289.92181 |
| | | | $C_{12}H_{6}{}^{35}Cl_{3}{}^{37}Cl$ | 291.91941 | 291.91886 |
| | | | $C_{12}H_{6}{}^{35}Cl_{2}{}^{37}Cl_{2}$ | 293.91646 | 293.91591 |
| | | | $C_{12}H_{6}{}^{35}Cl^{37}Cl_{3}$ | 295.91351 | 295.91296 |
| | | | $C_{12}H_{6}{}^{37}Cl_{4}$ | 297.91056 | 297.91001 |
| | | | | | |
| PCB-101 | 20.58 | $C_{12}H_{5}Cl_{5}$ | $C_{12}H_{5}{}^{35}Cl_{5}$ | 323.88338 | 323.88283 |
| | | | $C_{12}H_{5}{}^{35}Cl_{4}{}^{37}Cl$ | 325.88044 | 325.87989 |

| Compound | Retention time (min) | Formula | Isotopologue formula | Exact molecular weight (u) | Exact *m/z* value (u) |
|---|---|---|---|---|---|
| | | | $C_{12}H_5{}^{35}Cl_3{}^{37}Cl_2$ | 327.87749 | 327.87694 |
| | | | $C_{12}H_5{}^{35}Cl_2{}^{37}Cl_3$ | 329.87454 | 329.87399 |
| | | | $C_{12}H_5{}^{35}Cl^{37}Cl_4$ | 331.87159 | 331.87104 |
| | | | $C_{12}H_5{}^{37}Cl_5$ | 333.86864 | 333.86809 |
| | | | | | |
| PCB-138 | 27.54 | $C_{12}H_4Cl_6$ | $C_{12}H_4{}^{35}Cl_6$ | 357.84441 | 357.84386 |
| | | | $C_{12}H_4{}^{35}Cl_5{}^{37}Cl$ | 359.84146 | 359.84091 |
| | | | $C_{12}H_4{}^{35}Cl_4{}^{37}Cl_2$ | 361.83851 | 361.83796 |
| | | | $C_{12}H_4{}^{35}Cl_3{}^{37}Cl_3$ | 363.83556 | 363.83501 |
| | | | $C_{12}H_4{}^{35}Cl_2{}^{37}Cl_4$ | 365.83262 | 365.83207 |
| | | | $C_{12}H_4{}^{35}Cl^{37}Cl_5$ | 367.82967 | 367.82912 |
| | | | $C_{12}H_4{}^{37}Cl_6$ | 369.82672 | 369.82617 |
| | | | | | |
| PCB-153 | 30.02 | $C_{12}H_4Cl_6$ | Refer to PCB-138 | | |
| | | | | | |
| PCB-180 | 36.75 | $C_{12}H_3Cl_7$ | $C_{12}H_3{}^{35}Cl_7$ | 391.80544 | 391.80489 |
| | | | $C_{12}H_3{}^{35}Cl_6{}^{37}Cl$ | 393.80249 | 393.80194 |
| | | | $C_{12}H_3{}^{35}Cl_5{}^{37}Cl_2$ | 395.79954 | 395.79899 |
| | | | $C_{12}H_3{}^{35}Cl_4{}^{37}Cl_3$ | 397.79659 | 397.79604 |
| | | | $C_{12}H_3{}^{35}Cl_3{}^{37}Cl_4$ | 399.79364 | 399.79309 |
| | | | $C_{12}H_3{}^{35}Cl_2{}^{37}Cl_5$ | 401.79069 | 401.79014 |
| | | | $C_{12}H_3{}^{35}Cl^{37}Cl_6$ | 403.78774 | 403.78719 |
| | | | $C_{12}H_3{}^{37}Cl_7$ | 405.78479 | 405.78424 |
| | | | | | |
| Penta-CDD | 38.85 | $C_{12}H_3Cl_5O_2$ | $C_{12}H_3{}^{35}Cl_5O_2$ | 353.85756 | 353.85701 |
| | | | $C_{12}H_3{}^{35}Cl_4{}^{37}ClO_2$ | 355.85462 | 355.85407 |
| | | | $C_{12}H_3{}^{35}Cl_3{}^{37}Cl_2O_2$ | 357.85167 | 357.85112 |
| | | | $C_{12}H_3{}^{35}Cl_2{}^{37}Cl_3O_2$ | 359.84872 | 359.84817 |
| | | | $C_{12}H_3{}^{35}Cl^{37}Cl_4O_2$ | 361.84577 | 361.84522 |
| | | | $C_{12}H_3{}^{37}Cl_5O_2$ | 363.84282 | 363.84227 |
| | | | | | |
| Hexa-CDD-1 | 43.79 | $C_{12}H_2Cl_6O_2$ | $C_{12}H_2{}^{35}Cl_6O_2$ | 387.81859 | 387.81804 |
| | | | $C_{12}H_2{}^{35}Cl_5{}^{37}ClO_2$ | 389.81564 | 389.81509 |
| | | | $C_{12}H_2{}^{35}Cl_4{}^{37}Cl_2O_2$ | 391.81269 | 391.81214 |
| | | | $C_{12}H_2{}^{35}Cl_3{}^{37}Cl_3O_2$ | 393.80974 | 393.80919 |
| | | | $C_{12}H_2{}^{35}Cl_2{}^{37}Cl_4O_2$ | 395.80680 | 395.80625 |
| | | | $C_{12}H_2{}^{35}Cl^{37}Cl_5O_2$ | 397.80385 | 397.80330 |
| | | | $C_{12}H_2{}^{37}Cl_6O_2$ | 399.80090 | 399.80035 |
| | | | | | |
| Hexa-CDD-2 | 43.93 | $C_{12}H_2Cl_6O_2$ | Refer to Hexa-CDD-1 | | |
| Hexa-CDD-3 | 44.27 | $C_{12}H_2Cl_6O_2$ | Refer to Hexa-CDD-1 | | |
| | | | | | |
| Hepta-CDD | 47.86 | $C_{12}HC_{l7}O_2$ | $C_{12}H^{35}Cl_7O_2$ | 421.77962 | 421.77907 |
| | | | $C_{12}H^{35}Cl_6{}^{37}ClO_2$ | 423.77667 | 423.77612 |

| Compound | Retention time (min) | Formula | Isotopologue formula | Exact molecular weight (u) | Exact *m/z* value (u) |
|---|---|---|---|---|---|
| | | | $C_{12}H^{35}Cl_5{}^{37}Cl_2O_2$ | 425.77372 | 425.77317 |
| | | | $C_{12}H^{35}Cl_4{}^{37}Cl_3O_2$ | 427.77077 | 427.77022 |
| | | | $C_{12}H^{35}Cl_3{}^{37}Cl_4O_2$ | 429.76782 | 429.76727 |
| | | | $C_{12}H^{35}Cl_2{}^{37}Cl_5O_2$ | 431.76487 | 431.76432 |
| | | | $C_{12}H^{35}Cl^{37}Cl_6O_2$ | 433.76192 | 433.76137 |
| | | | $C_{12}H^{37}Cl_7O_2$ | 435.75897 | 435.75842 |
| | | | | | |
| OCDD | 51.06 | $C_{12}Cl_8O_2$ | $C_{12}{}^{35}Cl_8O_2$ | 455.74065 | 455.74010 |
| | | | $C_{12}{}^{35}Cl_7{}^{37}ClO_2$ | 457.73770 | 457.73715 |
| | | | $C_{12}{}^{35}Cl_6{}^{37}Cl_2O_2$ | 459.73475 | 459.73420 |
| | | | $C_{12}{}^{35}Cl_5{}^{37}Cl_3O_2$ | 461.73180 | 461.73125 |
| | | | $C_{12}{}^{35}Cl_4{}^{37}Cl_4O_2$ | 463.72885 | 463.72830 |
| | | | $C_{12}{}^{35}Cl_3{}^{37}Cl_5O_2$ | 465.72590 | 465.72535 |
| | | | $C_{12}{}^{35}Cl_2{}^{37}Cl_6O_2$ | 467.72295 | 467.72240 |
| | | | $C_{12}{}^{35}Cl^{37}Cl_7O_2$ | 469.72000 | 469.71945 |
| | | | $C_{12}{}^{37}Cl_8O_2$ | 471.71705 | 471.71650 |
| | | | | | |
| TCDF | 29.65 | $C_{12}H_4Cl_4O$ | $C_{12}H_4{}^{35}Cl_4O$ | 303.90163 | 303.90108 |
| | | | $C_{12}H_4{}^{35}Cl_3{}^{37}ClO$ | 305.89868 | 305.89813 |
| | | | $C_{12}H_4{}^{35}Cl_2{}^{37}Cl_2O$ | 307.89573 | 307.89518 |
| | | | $C_{12}H_4{}^{35}Cl^{37}Cl_3O$ | 309.89278 | 309.89223 |
| | | | $C_{12}H_4{}^{37}Cl_4O$ | 311.88983 | 311.88928 |
| | | | | | |
| Penta-CDF-1 | 36.86 | $C_{12}H_3Cl_5O$ | $C_{12}H_3{}^{35}Cl_5O$ | 337.86265 | 337.86210 |
| | | | $C_{12}H_3{}^{35}Cl_4{}^{37}ClO$ | 339.85971 | 339.85916 |
| | | | $C_{12}H_3{}^{35}Cl_3{}^{37}Cl_2O$ | 341.85676 | 341.85621 |
| | | | $C_{12}H_3{}^{35}Cl_2{}^{37}Cl_3O$ | 343.85381 | 343.85326 |
| | | | $C_{12}H_3{}^{35}Cl^{37}Cl_4O$ | 345.85086 | 345.85031 |
| | | | $C_{12}H_3{}^{37}Cl_5O$ | 347.84791 | 347.84736 |
| | | | | | |
| Penta-CDF-2 | 38.41 | $C_{12}H_3Cl_5O$ | Refer to Penta-CDF-1 | | |
| | | | | | |
| Hexa-CDF-1 | 42.64 | $C_{12}H_2Cl_6O$ | $C_{12}H_2{}^{35}Cl_6O$ | 371.82368 | 371.82313 |
| | | | $C_{12}H_2{}^{35}Cl_5{}^{37}ClO$ | 373.82073 | 373.82018 |
| | | | $C_{12}H_2{}^{35}Cl_4{}^{37}Cl_2O$ | 375.81778 | 375.81723 |
| | | | $C_{12}H_2{}^{35}Cl_3{}^{37}Cl_3O$ | 377.81483 | 377.81428 |
| | | | $C_{12}H_2{}^{35}Cl_2{}^{37}Cl_4O$ | 379.81189 | 379.81134 |
| | | | $C_{12}H_2{}^{35}Cl^{37}Cl_5O$ | 381.80894 | 381.80839 |
| | | | $C_{12}H_2{}^{37}Cl_6O$ | 383.80599 | 383.80544 |
| | | | | | |
| Hexa-CDF-2 | 42.81 | $C_{12}H_2Cl_6O$ | Refer to Hexa-CDF-1 | | |
| Hexa-CDF-3 | 43.61 | $C_{12}H_2Cl_6O$ | Refer to Hexa-CDF-1 | | |
| Hexa-CDF-4 | 44.71 | $C_{12}H_2Cl_6O$ | Refer to Hexa-CDF-1 | | |
| | | | | | |

| Compound | Retention time (min) | Formula | Isotopologue formula | Exact molecular weight (u) | Exact *m/z* value (u) |
|---|---|---|---|---|---|
| Hepta-CDF-1 | 46.58 | $C_{12}HCl_7O$ | $C_{12}H^{35}Cl_7O$ | 405.78471 | 405.78416 |
| | | | $C_{12}H^{35}Cl_6{}^{37}ClO$ | 407.78176 | 407.78121 |
| | | | $C_{12}H^{35}Cl_5{}^{37}Cl_2O$ | 409.77881 | 409.77826 |
| | | | $C_{12}H^{35}Cl_4{}^{37}Cl_3O$ | 411.77586 | 411.77531 |
| | | | $C_{12}H^{35}Cl_3{}^{37}Cl_4O$ | 413.77291 | 413.77236 |
| | | | $C_{12}H^{35}Cl_2{}^{37}Cl_5O$ | 415.76996 | 415.76941 |
| | | | $C_{12}H^{35}Cl^{37}Cl_6O$ | 417.76701 | 417.76646 |
| | | | $C_{12}H^{37}Cl_7O$ | 419.76406 | 419.76351 |
| | | | | | |
| Hepta-CDF-2 | 48.48 | $C_{12}HCl_7O$ | Refer to Hepta-CDF-1 | | |
| | | | | | |
| OCDF | 51.27 | $C_{12}Cl_8O$ | $C_{12}{}^{35}Cl_8O$ | 441.74279 | 441.74224 |
| | | | $C_{12}{}^{35}Cl_7{}^{37}ClO$ | 443.73984 | 443.73929 |
| | | | $C_{12}{}^{35}Cl_6{}^{37}Cl_2O$ | 445.73689 | 445.73634 |
| | | | $C_{12}{}^{35}Cl_5{}^{37}Cl_3O$ | 447.73394 | 447.73339 |
| | | | $C_{12}{}^{35}Cl_4{}^{37}Cl_4O$ | 439.74574 | 439.74519 |
| | | | $C_{12}{}^{35}Cl_3{}^{37}Cl_5O$ | 449.73099 | 449.73044 |
| | | | $C_{12}{}^{35}Cl_2{}^{37}Cl_6O$ | 451.72804 | 451.72749 |
| | | | $C_{12}{}^{35}Cl^{37}Cl_7O$ | 453.72509 | 453.72454 |
| | | | $C_{12}{}^{37}Cl_8O_2$ | 455.72214 | 455.72159 |

**Table S-3.** Isotope ratios, delta values ($\delta^hE$, referenced to SMOE) and relative variations ($\Delta^hE$) derived from different retention time segments of the investigated compounds. SMOE: Standard Mean Ocean Element.

| Compound | Retention time segment | Isotope ratio (mean, n=5) | SD (1σ, ‰) | RSD (‰) | $\delta^hE$ (vs SMOE, ‰) | SD (1σ, ‰) | $\Delta^hE$ (mean, n=5, ‰) | SD (1σ, ‰) |
|---|---|---|---|---|---|---|---|---|
| $^{13}C_6$-HBB | Overall | 0.97090 | 0.38 | 0.39 | -1.93 | 0.39 | 0.00 | 0.39 |
| | T1 | 0.98074 | 6.32 | 6.44 | 8.19 | 6.49 | 10.13 | 6.51 |
| | T2 | 0.97446 | 1.39 | 1.43 | 1.74 | 1.43 | 3.67 | 1.43 |
| | T3 | 0.97043 | 0.60 | 0.62 | -2.41 | 0.62 | -0.49 | 0.62 |
| | T4 | 0.96678 | 1.97 | 2.03 | -6.17 | 2.02 | -4.25 | 2.03 |
| | T5 | 0.96181 | 0.96 | 1.00 | -11.27 | 0.99 | -9.37 | 0.99 |
| | | | | | | | | |
| HBB | Overall | 0.97189 | 0.34 | 0.35 | -0.91 | 0.35 | 0.00 | 0.35 |
| | T1 | 0.97902 | 3.51 | 3.58 | 6.42 | 3.61 | 7.34 | 3.61 |
| | T2 | 0.97549 | 0.82 | 0.84 | 2.79 | 0.84 | 3.71 | 0.84 |
| | T3 | 0.97220 | 0.76 | 0.78 | -0.59 | 0.78 | 0.32 | 0.78 |
| | T4 | 0.96833 | 0.63 | 0.65 | -4.57 | 0.65 | -3.67 | 0.65 |
| | T5 | 0.96336 | 4.21 | 4.37 | -9.67 | 4.33 | -8.77 | 4.33 |
| | | | | | | | | |
| BDE-77 | Overall | 0.97287 | 0.47 | 0.48 | 0.09 | 0.48 | 0.00 | 0.48 |
| | T1 | 0.99270 | 3.43 | 3.45 | 20.48 | 3.52 | 20.39 | 3.52 |
| | T2 | 0.97805 | 1.87 | 1.91 | 5.42 | 1.93 | 5.33 | 1.92 |
| | T3 | 0.97053 | 1.36 | 1.40 | -2.30 | 1.40 | -2.40 | 1.40 |
| | T4 | 0.96506 | 2.31 | 2.39 | -7.93 | 2.37 | -8.02 | 2.37 |
| | T5 | 0.95570 | 5.28 | 5.52 | 1.20 | 5.43 | -17.64 | 5.42 |
| | | | | | | | | |
| OBDD | Overall | 0.97545 | 1.74 | 1.78 | 2.74 | 1.79 | 0.00 | 1.78 |
| | T1 | 0.98900 | 5.77 | 5.84 | 16.68 | 5.94 | 13.89 | 5.92 |
| | T2 | 0.97631 | 5.83 | 5.98 | 3.64 | 6.00 | 0.89 | 5.98 |
| | T3 | 0.97302 | 4.74 | 4.88 | 0.25 | 4.88 | -2.49 | 4.86 |
| | T4 | 0.96937 | 5.17 | 5.34 | -3.50 | 5.32 | -6.22 | 5.30 |
| | T5 | 0.97394 | 4.55 | 4.67 | 1.20 | 4.68 | -1.54 | 4.67 |
| | | | | | | | | |
| HCB | Overall | 0.31665 | 0.39 | 1.22 | -9.25 | 1.21 | 0.00 | 1.22 |
| | T1 | 0.32580 | 1.31 | 4.02 | 19.35 | 4.10 | 28.87 | 4.14 |
| | T2 | 0.31918 | 0.65 | 2.02 | -1.34 | 2.02 | 7.99 | 2.04 |
| | T3 | 0.31516 | 0.64 | 2.03 | -13.92 | 2.00 | -4.71 | 2.02 |
| | T4 | 0.31354 | 0.65 | 2.07 | -18.98 | 2.03 | -9.82 | 2.05 |
| | T5 | 0.30954 | 0.40 | 1.29 | -31.50 | 1.25 | -22.45 | 1.26 |
| | | | | | | | | |

| Compound | Retention time segment | Isotope ratio (mean, n=5) | SD (1σ, ‰) | RSD (‰) | $\delta^hE$ (vs SMOE, ‰) | SD (1σ, ‰) | $\Delta^hE$ (mean, n=5, ‰) | SD (1σ, ‰) |
|---|---|---|---|---|---|---|---|---|
| Me-TCS | Overall | 0.34310 | 0.31 | 0.92 | 73.51 | 0.98 | 0.00 | 0.92 |
| | T1 | 0.34578 | 1.73 | 5.01 | 81.88 | 5.42 | 7.80 | 5.05 |
| | T2 | 0.34509 | 1.17 | 3.38 | 79.72 | 3.65 | 5.79 | 3.40 |
| | T3 | 0.34293 | 0.73 | 2.12 | 72.95 | 2.27 | -0.51 | 2.12 |
| | T4 | 0.34113 | 0.88 | 2.57 | 67.34 | 2.74 | -5.74 | 2.55 |
| | T5 | 0.33124 | 0.51 | 1.53 | 36.40 | 1.58 | -34.56 | 1.47 |
| | | | | | | | | |
| o,p-DDE | Overall | 0.31939 | 1.33 | 4.18 | -0.68 | 4.18 | 0.00 | 4.18 |
| | T1 | 0.32545 | 1.80 | 5.54 | 18.27 | 5.64 | 18.96 | 5.64 |
| | T2 | 0.32350 | 1.25 | 3.87 | 12.16 | 3.92 | 12.85 | 3.92 |
| | T3 | 0.32081 | 1.27 | 3.95 | 3.75 | 3.97 | 4.43 | 3.97 |
| | T4 | 0.31533 | 0.53 | 1.69 | -13.39 | 1.66 | -12.72 | 1.67 |
| | T5 | 0.30940 | 0.09 | 0.29 | -31.94 | 0.28 | -31.28 | 0.28 |
| | | | | | | | | |
| p,p-DDE | Overall | 0.31843 | 0.88 | 2.75 | -3.68 | 2.74 | 0.00 | 2.75 |
| | T1 | 0.32174 | 1.34 | 4.16 | 6.67 | 4.19 | 10.38 | 4.20 |
| | T2 | 0.32065 | 1.40 | 4.35 | 3.25 | 4.37 | 6.95 | 4.38 |
| | T3 | 0.31914 | 0.91 | 2.86 | -1.47 | 2.86 | 2.21 | 2.87 |
| | T4 | 0.31648 | 1.00 | 3.14 | -9.80 | 3.11 | -6.14 | 3.13 |
| | T5 | 0.31247 | 1.75 | 5.59 | -22.34 | 5.46 | -18.73 | 5.48 |
| | | | | | | | | |
| o,p-DDD | Overall | 0.31570 | 0.88 | 2.78 | -12.24 | 2.75 | 0.00 | 2.78 |
| | T1 | 0.32651 | 2.27 | 6.96 | 21.58 | 7.11 | 34.25 | 7.20 |
| | T2 | 0.31704 | 0.97 | 3.07 | -8.05 | 3.05 | 4.24 | 3.09 |
| | T3 | 0.31650 | 1.09 | 3.44 | -9.74 | 3.41 | 2.54 | 3.45 |
| | T4 | 0.31415 | 0.22 | 0.69 | -17.07 | 0.67 | -4.89 | 0.68 |
| | T5 | 0.30948 | 0.74 | 2.38 | -31.68 | 2.30 | -19.67 | 2.33 |
| | | | | | | | | |
| p,p-DDD | Overall | 0.31590 | 2.22 | 7.02 | -11.59 | 6.94 | 0.00 | 7.02 |
| | T1 | 0.32179 | 3.51 | 10.92 | 6.82 | 11.00 | 18.63 | 11.12 |
| | T2 | 0.31420 | 1.21 | 3.85 | -16.93 | 3.78 | -5.40 | 3.83 |
| | T3 | 0.31494 | 4.42 | 14.03 | -14.62 | 13.82 | -3.06 | 13.99 |
| | | | | | | | | |
| | | | | | | | | |
| | | | | | | | | |
| o,p-DDT | Overall | 0.31460 | 1.66 | 5.28 | -15.66 | 5.20 | 0.00 | 5.28 |
| | T1 | 0.31820 | 2.89 | 9.09 | -4.39 | 9.05 | 11.44 | 9.20 |
| | T2 | 0.31460 | 1.86 | 5.91 | -15.68 | 5.82 | -0.02 | 5.91 |
| | T3 | 0.30672 | 3.79 | 12.37 | -40.32 | 11.87 | -25.05 | 12.06 |
| | | | | | | | | |

| Compound | Retention time segment | Isotope ratio (mean, n=5) | SD (1σ, ‰) | RSD (‰) | $\delta^hE$ (vs SMOE, ‰) | SD (1σ, ‰) | $\Delta^hE$ (mean, n=5, ‰) | SD (1σ, ‰) |
|---|---|---|---|---|---|---|---|---|
| | | | | | | | | |
| | | | | | | | | |
| p,p-DDT | Overall | 0.31569 | 1.29 | 4.10 | -12.28 | 4.05 | 0.00 | 4.10 |
| | T1 | 0.32555 | 6.13 | 18.84 | 18.59 | 19.19 | 31.24 | 19.43 |
| | T2 | 0.31703 | 2.78 | 8.77 | -8.06 | 8.70 | 4.27 | 8.81 |
| | T3 | 0.31339 | 1.97 | 6.27 | -19.47 | 6.15 | -7.28 | 6.23 |
| | T4 | 0.31425 | 0.30 | 0.96 | -16.78 | 0.94 | -4.56 | 0.95 |
| | | | | | | | | |
| | | | | | | | | |
| PCB-18 | Overall | 0.34158 | 0.42 | 1.22 | 68.75 | 1.30 | 0.00 | 1.22 |
| | T1 | 0.35378 | 1.35 | 3.81 | 106.91 | 4.22 | 35.70 | 3.95 |
| | T2 | 0.34860 | 1.09 | 3.12 | 90.69 | 3.40 | 20.53 | 3.18 |
| | T3 | 0.34229 | 1.72 | 5.02 | 70.98 | 5.38 | 2.08 | 5.03 |
| | T4 | 0.33575 | 1.25 | 3.72 | 50.50 | 3.91 | -17.07 | 3.66 |
| | T5 | 0.32968 | 1.72 | 5.20 | 31.51 | 5.37 | -34.85 | 5.02 |
| | | | | | | | | |
| PCB-28 | Overall | 0.34156 | 0.89 | 2.61 | 68.69 | 2.79 | 0.00 | 2.61 |
| | T1 | 0.34895 | 2.01 | 5.77 | 91.79 | 6.30 | 21.61 | 5.89 |
| | T2 | 0.34538 | 1.64 | 4.75 | 80.64 | 5.13 | 11.19 | 4.80 |
| | T3 | 0.34188 | 1.18 | 3.46 | 69.69 | 3.70 | 0.94 | 3.46 |
| | T4 | 0.33740 | 0.76 | 2.24 | 55.68 | 2.37 | -12.18 | 2.22 |
| | T5 | 0.33343 | 0.43 | 1.28 | 43.24 | 1.34 | -23.81 | 1.25 |
| | | | | | | | | |
| PCB-52 | Overall | 0.32930 | 0.26 | 0.80 | 30.32 | 0.83 | 0.00 | 0.80 |
| | T1 | 0.33393 | 2.00 | 5.99 | 44.81 | 6.26 | 14.06 | 6.08 |
| | T2 | 0.33208 | 0.65 | 1.94 | 39.02 | 2.02 | 8.44 | 1.96 |
| | T3 | 0.32925 | 0.20 | 0.59 | 30.17 | 0.61 | -0.15 | 0.59 |
| | T4 | 0.32555 | 0.85 | 2.60 | 18.58 | 2.65 | -11.39 | 2.57 |
| | T5 | 0.32048 | 2.30 | 7.18 | 2.74 | 7.20 | -26.77 | 6.99 |
| | | | | | | | | |
| PCB-101 | Overall | 0.32144 | 1.08 | 3.36 | 5.72 | 3.38 | 0.00 | 3.36 |
| | T1 | 0.31991 | 1.28 | 4.00 | 0.96 | 4.00 | -4.73 | 3.98 |
| | T2 | 0.32195 | 1.17 | 3.65 | 7.33 | 3.67 | 1.60 | 3.65 |
| | T3 | 0.32168 | 0.96 | 2.99 | 6.47 | 3.01 | 0.74 | 2.99 |
| | T4 | 0.32107 | 1.34 | 4.19 | 4.58 | 4.20 | -1.13 | 4.18 |
| | T5 | 0.31892 | 1.69 | 5.31 | -2.15 | 5.30 | -7.82 | 5.27 |
| | | | | | | | | |
| PCB-138 | Overall | 0.32640 | 0.45 | 1.37 | 21.25 | 1.40 | 0.00 | 1.37 |
| | T1 | 0.32314 | 2.20 | 6.80 | 11.06 | 6.87 | -9.98 | 6.73 |
| | T2 | 0.32578 | 1.08 | 3.32 | 19.30 | 3.38 | -1.92 | 3.31 |

| Compound | Retention time segment | Isotope ratio (mean, n=5) | SD (1σ, ‰) | RSD (‰) | $\delta^hE$ (vs SMOE, ‰) | SD (1σ, ‰) | $\Delta^hE$ (mean, n=5, ‰) | SD (1σ, ‰) |
|---|---|---|---|---|---|---|---|---|
| | T3 | 0.32646 | 0.56 | 1.71 | 21.44 | 1.75 | 0.18 | 1.71 |
| | T4 | 0.32714 | 0.33 | 1.00 | 23.55 | 1.02 | 2.25 | 1.00 |
| | T5 | 0.32540 | 1.73 | 5.32 | 18.12 | 5.42 | -3.06 | 5.31 |
| | | | | | | | | |
| PCB-153 | Overall | 0.32717 | 0.26 | 0.80 | 23.67 | 0.82 | 0.00 | 0.80 |
| | T1 | 0.32456 | 0.63 | 1.93 | 15.49 | 1.96 | -7.99 | 1.91 |
| | T2 | 0.32658 | 0.76 | 2.33 | 21.82 | 2.38 | -1.80 | 2.32 |
| | T3 | 0.32732 | 0.35 | 1.08 | 24.11 | 1.11 | 0.43 | 1.08 |
| | T4 | 0.32768 | 0.32 | 0.98 | 25.26 | 1.00 | 1.56 | 0.98 |
| | T5 | 0.32883 | 2.56 | 7.78 | 28.85 | 8.01 | 5.06 | 7.82 |
| | | | | | | | | |
| PCB-180 | Overall | 0.33067 | 0.97 | 2.94 | 34.60 | 3.05 | 0.00 | 2.94 |
| | T1 | 0.32955 | 1.56 | 4.74 | 31.11 | 4.89 | -3.37 | 4.73 |
| | T2 | 0.33037 | 0.94 | 2.84 | 33.66 | 2.94 | -0.91 | 2.84 |
| | T3 | 0.33075 | 1.05 | 3.18 | 34.85 | 3.29 | 0.25 | 3.18 |
| | T4 | 0.33112 | 1.07 | 3.23 | 36.01 | 3.35 | 1.37 | 3.23 |
| | T5 | 0.32937 | 1.42 | 4.31 | 30.55 | 4.44 | -3.91 | 4.29 |
| | | | | | | | | |
| Penta-CDD | Overall | 0.32111 | 0.44 | 1.38 | 4.70 | 1.39 | 0.00 | 1.38 |
| | T1 | 0.32573 | 1.09 | 3.35 | 19.16 | 3.42 | 14.39 | 3.40 |
| | T2 | 0.32264 | 0.13 | 0.40 | 9.50 | 0.41 | 4.77 | 0.40 |
| | T3 | 0.32074 | 0.80 | 2.51 | 3.52 | 2.51 | -1.17 | 2.50 |
| | T4 | 0.31916 | 0.50 | 1.56 | -1.41 | 1.56 | -6.08 | 1.55 |
| | T5 | 0.31909 | 1.93 | 6.04 | -1.64 | 6.03 | -6.31 | 6.00 |
| | | | | | | | | |
| Hexa-CDD-1 | Overall | 0.31638 | 0.23 | 0.71 | -10.11 | 0.70 | 0.00 | 0.71 |
| | T1 | 0.32292 | 2.00 | 6.20 | 10.35 | 6.27 | 20.67 | 6.33 |
| | T2 | 0.31981 | 0.93 | 2.90 | 0.63 | 2.90 | 10.85 | 2.93 |
| | T3 | 0.31722 | 0.56 | 1.75 | -7.48 | 1.74 | 2.66 | 1.75 |
| | T4 | 0.31382 | 0.44 | 1.39 | -18.10 | 1.37 | -8.08 | 1.38 |
| | T5 | 0.31168 | 1.96 | 6.28 | -24.82 | 6.12 | -14.86 | 6.18 |
| | | | | | | | | |
| Hexa-CDD-2 | Overall | 0.31789 | 0.35 | 1.10 | -5.38 | 1.09 | 0.00 | 1.10 |
| | T1 | 0.32217 | 0.78 | 2.41 | 8.03 | 2.43 | 13.48 | 2.45 |
| | T2 | 0.31986 | 0.80 | 2.50 | 0.78 | 2.50 | 6.20 | 2.51 |
| | T3 | 0.31691 | 0.47 | 1.49 | -8.44 | 1.48 | -3.08 | 1.48 |
| | T4 | 0.31362 | 0.95 | 3.04 | -18.75 | 2.98 | -13.44 | 3.00 |
| | T5 | 0.31346 | 2.32 | 7.39 | -19.23 | 7.24 | -13.92 | 7.28 |

| Compound | Retention time segment | Isotope ratio (mean, n=5) | SD (1σ, ‰) | RSD (‰) | $\delta^hE$ (vs SMOE, ‰) | SD (1σ, ‰) | $\Delta^hE$ (mean, n=5, ‰) | SD (1σ, ‰) |
|---|---|---|---|---|---|---|---|---|
| | | | | | | | | |
| Hexa-CDD-3 | Overall | 0.31810 | 0.08 | 0.26 | -4.71 | 0.26 | 0.00 | 0.26 |
| | T1 | 0.32232 | 0.67 | 2.09 | 8.49 | 2.10 | 13.27 | 2.11 |
| | T2 | 0.32160 | 0.64 | 2.00 | 6.22 | 2.01 | 10.98 | 2.02 |
| | T3 | 0.31833 | 0.20 | 0.62 | -4.02 | 0.61 | 0.70 | 0.62 |
| | T4 | 0.31410 | 0.67 | 2.14 | -17.23 | 2.10 | -12.57 | 2.11 |
| | T5 | 0.31150 | 1.93 | 6.21 | -25.39 | 6.05 | -20.77 | 6.08 |
| | | | | | | | | |
| Hepta-CDD | Overall | 0.31698 | 0.64 | 2.02 | -8.24 | 2.00 | 0.00 | 2.02 |
| | T1 | 0.32748 | 2.26 | 6.89 | 24.63 | 7.06 | 33.14 | 7.12 |
| | T2 | 0.32076 | 1.28 | 3.99 | 3.61 | 4.01 | 11.95 | 4.04 |
| | T3 | 0.31641 | 0.71 | 2.23 | -10.00 | 2.21 | -1.78 | 2.23 |
| | T4 | 0.31187 | 0.69 | 2.21 | -24.22 | 2.15 | -16.11 | 2.17 |
| | T5 | 0.31227 | 2.22 | 7.11 | -22.95 | 6.95 | -14.83 | 7.01 |
| | | | | | | | | |
| OCDD | Overall | 0.31910 | 0.51 | 1.61 | -1.59 | 1.60 | 0.00 | 1.61 |
| | T1 | 0.32990 | 2.96 | 8.97 | 32.19 | 9.25 | 33.83 | 9.27 |
| | T2 | 0.32269 | 0.83 | 2.57 | 9.65 | 2.59 | 11.26 | 2.60 |
| | T3 | 0.31763 | 0.96 | 3.01 | -6.20 | 2.99 | -4.61 | 3.00 |
| | T4 | 0.31239 | 0.75 | 2.39 | -22.60 | 2.34 | -21.04 | 2.34 |
| | T5 | 0.31039 | 1.88 | 6.05 | -28.84 | 5.87 | -27.29 | 5.88 |
| | | | | | | | | |
| TCDF | Overall | 0.32180 | 0.89 | 2.76 | 6.87 | 2.78 | 0.00 | 2.76 |
| | T1 | 0.32149 | 2.88 | 8.97 | 5.89 | 9.02 | -0.98 | 8.96 |
| | T2 | 0.32170 | 1.26 | 3.92 | 6.53 | 3.94 | -0.34 | 3.92 |
| | T3 | 0.32210 | 1.00 | 3.11 | 7.78 | 3.13 | 0.90 | 3.11 |
| | T4 | 0.32035 | 0.77 | 2.41 | 2.33 | 2.42 | -4.51 | 2.40 |
| | T5 | 0.32273 | 2.05 | 6.36 | 9.75 | 6.42 | 2.87 | 6.38 |
| | | | | | | | | |
| Penta-CDF-1 | Overall | 0.31945 | 0.20 | 0.61 | -0.49 | 0.61 | 0.00 | 0.61 |
| | T1 | 0.31913 | 1.19 | 3.73 | -1.50 | 3.72 | -1.01 | 3.73 |
| | T2 | 0.31981 | 0.50 | 1.55 | 0.64 | 1.56 | 1.13 | 1.56 |
| | T3 | 0.31944 | 0.22 | 0.70 | -0.54 | 0.70 | -0.05 | 0.70 |
| | T4 | 0.31825 | 0.65 | 2.06 | -4.25 | 2.05 | -3.76 | 2.05 |
| | T5 | 0.31895 | 2.25 | 7.07 | -2.07 | 7.05 | -1.58 | 7.05 |
| | | | | | | | | |
| Penta-CDF-2 | Overall | 0.32103 | 0.24 | 0.74 | 4.44 | 0.75 | 0.00 | 0.74 |

| Compound | Retention time segment | Isotope ratio (mean, n=5) | SD (1σ, ‰) | RSD (‰) | $\delta^hE$ (vs SMOE, ‰) | SD (1σ, ‰) | $\Delta^hE$ (mean, n=5, ‰) | SD (1σ, ‰) |
|---|---|---|---|---|---|---|---|---|
| | T1 | 0.32257 | 0.72 | 2.23 | 9.25 | 2.25 | 4.80 | 2.24 |
| | T2 | 0.32192 | 0.59 | 1.84 | 7.23 | 1.85 | 2.78 | 1.84 |
| | T3 | 0.32121 | 0.18 | 0.56 | 5.00 | 0.56 | 0.56 | 0.56 |
| | T4 | 0.31927 | 0.51 | 1.61 | -1.06 | 1.61 | -5.48 | 1.60 |
| | T5 | 0.31853 | 1.35 | 4.23 | -3.38 | 4.22 | -7.79 | 4.20 |
| | | | | | | | | |
| Hexa-CDF-1 | Overall | 0.31886 | 0.37 | 1.15 | -2.34 | 1.14 | 0.00 | 1.15 |
| | T1 | 0.32444 | 1.63 | 5.02 | 15.11 | 5.09 | 17.49 | 5.11 |
| | T2 | 0.32192 | 0.42 | 1.31 | 7.22 | 1.32 | 9.58 | 1.32 |
| | T3 | 0.31972 | 0.46 | 1.44 | 0.34 | 1.44 | 2.68 | 1.44 |
| | T4 | 0.31589 | 0.66 | 2.09 | -11.64 | 2.07 | -9.33 | 2.07 |
| | T5 | 0.31272 | 0.64 | 2.04 | -21.56 | 2.00 | -19.27 | 2.00 |
| | | | | | | | | |
| Hexa-CDF-2 | Overall | 0.31930 | 0.31 | 0.98 | -0.98 | 0.98 | 0.00 | 0.98 |
| | T1 | 0.32482 | 1.53 | 4.71 | 16.31 | 4.79 | 17.30 | 4.79 |
| | T2 | 0.32223 | 0.73 | 2.26 | 8.21 | 2.28 | 9.20 | 2.28 |
| | T3 | 0.31890 | 0.56 | 1.76 | -2.21 | 1.75 | -1.24 | 1.76 |
| | T4 | 0.31602 | 0.81 | 2.56 | -11.22 | 2.53 | -10.26 | 2.53 |
| | T5 | 0.31298 | 0.82 | 2.62 | -20.73 | 2.56 | -19.77 | 2.57 |
| | | | | | | | | |
| Hexa-CDF-3 | Overall | 0.31995 | 0.12 | 0.38 | 1.08 | 0.38 | 0.00 | 0.38 |
| | T1 | 0.32764 | 0.99 | 3.01 | 25.14 | 3.09 | 24.04 | 3.08 |
| | T2 | 0.32340 | 0.65 | 2.02 | 11.86 | 2.04 | 10.77 | 2.04 |
| | T3 | 0.31979 | 0.58 | 1.80 | 0.56 | 1.80 | -0.52 | 1.80 |
| | T4 | 0.31589 | 0.53 | 1.68 | -11.63 | 1.66 | -12.70 | 1.66 |
| | T5 | 0.31250 | 1.79 | 5.73 | -22.25 | 5.60 | -23.30 | 5.59 |
| | | | | | | | | |
| Hexa-CDF-4 | Overall | 0.31792 | 0.30 | 0.95 | -5.27 | 0.94 | 0.00 | 0.95 |
| | T1 | 0.32597 | 0.34 | 1.05 | 19.92 | 1.07 | 25.32 | 1.07 |
| | T2 | 0.32142 | 0.81 | 2.51 | 5.66 | 2.52 | 10.99 | 2.54 |
| | T3 | 0.31736 | 0.56 | 1.78 | -7.04 | 1.76 | -1.78 | 1.77 |
| | T4 | 0.31305 | 0.68 | 2.17 | -20.52 | 2.13 | -15.32 | 2.14 |
| | T5 | 0.31093 | 1.78 | 5.72 | -27.14 | 5.56 | -21.98 | 5.59 |
| | | | | | | | | |
| Hepta-CDF-1 | Overall | 0.31973 | 0.31 | 0.97 | 0.39 | 0.97 | 0.00 | 0.97 |
| | T1 | 0.32593 | 1.50 | 4.59 | 19.79 | 4.68 | 19.39 | 4.68 |

| Compound | Retention time segment | Isotope ratio (mean, n=5) | SD (1σ, ‰) | RSD (‰) | $\delta^hE$ (vs SMOE, ‰) | SD (1σ, ‰) | $\Delta^hE$ (mean, n=5, ‰) | SD (1σ, ‰) |
|---|---|---|---|---|---|---|---|---|
| | T2 | 0.32275 | 0.45 | 1.40 | 9.82 | 1.42 | 9.43 | 1.42 |
| | T3 | 0.31897 | 0.39 | 1.24 | -1.99 | 1.23 | -2.38 | 1.23 |
| | T4 | 0.31564 | 0.81 | 2.55 | -12.40 | 2.52 | -12.79 | 2.52 |
| | T5 | 0.31337 | 2.96 | 9.46 | -19.51 | 9.27 | -19.89 | 9.27 |
| | | | | | | | | |
| Hepta-CDF-2 | Overall | 0.31969 | 0.38 | 1.20 | 0.25 | 1.20 | -0.13 | 1.20 |
| | T1 | 0.32820 | 1.26 | 3.84 | 26.89 | 3.94 | 26.49 | 3.94 |
| | T2 | 0.32346 | 0.71 | 2.21 | 12.05 | 2.24 | 11.66 | 2.24 |
| | T3 | 0.31872 | 0.46 | 1.43 | -2.79 | 1.43 | -3.17 | 1.43 |
| | T4 | 0.31440 | 0.79 | 2.52 | -16.29 | 2.48 | -16.67 | 2.48 |
| | T5 | 0.31531 | 1.65 | 5.23 | -13.44 | 5.16 | -13.82 | 5.16 |
| | | | | | | | | |
| OCDF | Overall | 0.31330 | 0.49 | 1.58 | -19.73 | 1.55 | 0.00 | 1.58 |
| | T1 | 0.32236 | 1.11 | 3.45 | 8.59 | 3.48 | 28.89 | 3.55 |
| | T2 | 0.31692 | 0.95 | 3.00 | -8.40 | 2.97 | 11.56 | 3.03 |
| | T3 | 0.31186 | 1.08 | 3.48 | -24.26 | 3.39 | -4.62 | 3.46 |
| | T4 | 0.30726 | 1.11 | 3.62 | -38.65 | 3.48 | -19.30 | 3.55 |
| | T5 | 0.30802 | 2.86 | 9.30 | -36.27 | 8.96 | -16.88 | 9.14 |

**Table S-4.** Overall isotope ratios and $\delta^hE$ values (referenced to SMOE) and isotope fractionation extents ($\Delta'^hE$) of all the investigated compounds along with precision results of the developed CSIA method.

| Abbreviation | Ratio (mean, n=5) | SD (1σ, ‰) | RSD (‰) | $\delta^hE$ (vs SMOE, ‰) | SD (1σ, ‰) | $\Delta'^hE$ (mean, n=5, ‰) |
|---|---|---|---|---|---|---|
| $^{13}C_6$-HBB | 0.97090 | 0.38 | 0.39 | -1.93 | 0.39 | 19.69 |
| HBB | 0.97189 | 0.34 | 0.35 | -0.91 | 0.35 | 16.25 |
| BDE-77 | 0.97287 | 0.47 | 0.48 | 0.09 | 0.48 | 38.71 |
| OBDD | 0.97545 | 1.74 | 1.78 | 2.75 | 1.79 | 15.46 |
| HCB | 0.31665 | 0.39 | 1.22 | -9.25 | 1.21 | 52.51 |
| Me-TCS | 0.34310 | 0.31 | 0.92 | 73.51 | 0.98 | 43.88 |
| o,p'-DDE | 0.31939 | 1.34 | 4.18 | -0.68 | 4.18 | 51.87 |
| p,p'-DDE | 0.31843 | 0.88 | 2.75 | -3.68 | 2.74 | 29.67 |
| o,p'-DDD | 0.31570 | 0.88 | 2.78 | -12.24 | 2.75 | 55.00 |
| p,p'-DDD | 0.31590 | 2.22 | 7.02 | -11.59 | 6.94 | 21.76 |
| o,p'-DDT | 0.31460 | 1.66 | 5.28 | -15.66 | 5.20 | 37.43 |
| p,p'-DDT | 0.31569 | 1.29 | 4.10 | 73.51 | 4.05 | 35.97 |
| PCB-18 | 0.34158 | 0.42 | 1.21 | 68.75 | 1.30 | 73.09 |
| PCB-28 | 0.34156 | 0.89 | 2.61 | 68.69 | 2.79 | 46.54 |
| PCB-52 | 0.32930 | 0.26 | 0.80 | 30.32 | 0.83 | 41.96 |
| PCB-101 | 0.32144 | 1.08 | 3.36 | 5.72 | 3.38 | 3.11 |
| PCB-138 | 0.32640 | 0.45 | 1.38 | 21.25 | 1.40 | -6.94 |
| PCB-153 | 0.32718 | 0.26 | 0.80 | 23.67 | 0.82 | -12.98 |
| PCB-180 | 0.33067 | 0.97 | 2.94 | 34.60 | 3.05 | 0.54 |
| Penta-CDD | 0.32111 | 0.44 | 1.38 | 4.70 | 1.39 | 20.83 |
| Hexa-CDD-1 | 0.31638 | 0.23 | 0.71 | -10.11 | 0.71 | 36.07 |
| Hexa-CDD-2 | 0.31789 | 0.35 | 1.10 | -5.38 | 1.09 | 27.79 |
| Hexa-CDD-3 | 0.31810 | 0.08 | 0.26 | -4.71 | 0.26 | 34.76 |
| Hepta-CDD | 0.31698 | 0.64 | 2.02 | -8.24 | 2.00 | 48.69 |
| OCDD | 0.31910 | 0.51 | 1.61 | -1.59 | 1.61 | 62.84 |
| TCDF | 0.32180 | 0.89 | 2.76 | -19.73 | 6.87 | -3.83 |
| Penta-CDF-1 | 0.31945 | 0.20 | 0.61 | 4.44 | 0.61 | 0.57 |
| Penta-CDF-2 | 0.32103 | 0.24 | 0.74 | 4.44 | 0.75 | 12.68 |
| Hexa-CDF-1 | 0.31886 | 0.37 | 1.15 | -2.34 | 1.15 | 37.48 |
| Hexa-CDF-2 | 0.31930 | 0.31 | 0.98 | -0.98 | 0.98 | 37.83 |
| Hexa-CDF-3 | 0.31995 | 0.12 | 0.38 | 1.08 | 0.38 | 48.46 |
| Hexa-CDF-4 | 0.31792 | 0.30 | 0.95 | -5.27 | 0.94 | 48.37 |
| Hepta-CDF-1 | 0.31973 | 0.31 | 0.97 | 0.39 | 0.97 | 40.08 |
| Hepta-CDF-2 | 0.31969 | 0.38 | 1.20 | 0.26 | 1.20 | 40.88 |
| OCDF | 0.31330 | 0.50 | 1.58 | -19.73 | 1.55 | 46.56 |